\documentclass[]{aa_mine}

\usepackage{graphicx}
\usepackage{txfonts}

\graphicspath{{imagens/}}

\newlength{\picwd}
\newlength{\thinpic}
\usepackage[english]{babel}
\usepackage[utf8]{inputenc}
\usepackage{natbib}
\usepackage{datetime}
\usepackage{xcolor}

\newcommand{\e}[1]{\ensuremath{\times 10^{#1}}} %
\newcommand{\un}[1]{\ensuremath{\ \mathrm{#1}}}

\newcommand{\rot}[1]{\ensuremath{\nabla\times {#1}}}
\newcommand{\diver}[1]{\ensuremath{\nabla\cdot {#1}}}

\newcommand{\DD}{\mbox{\boldmath ${\cal D}$}}

\newcommand{\new}[1]{{#1}}

\begin{document}

\title{Soft X-ray emission in kink-unstable coronal loops}

\author{
  R. F. Pinto \inst{1,2}
  \and
  N. Vilmer \inst{1}
  \and 
  A. S. Brun \inst{2}
}

\institute{LESIA, Observatoire de Paris, CNRS, UPMC, Université Paris-Diderot, 5 place Jules Janssen, 92195 Meudon, France\\
  \email{rui.pinto@obspm.fr}
  \and
Laboratoire AIM Paris-Saclay, CEA/Irfu Universit\'e Paris-Diderot CNRS/ 
  INSU, 91191 Gif-sur-Yvette, France
}


 
  \abstract
  {
   Solar flares are associated with intense soft X-ray emission generated by the hot flaring plasma in coronal magnetic loops.
    Kink unstable twisted flux-ropes provide a source of  magnetic energy which can be released impulsively and account for the heating of the plasma in flares.
  }
  %
  {
    We investigate the temporal, spectral and spatial evolution of the properties of the thermal continuum X-ray emission produced in such kink-unstable magnetic flux-ropes and we discuss the results of the simulations with respect to solar flare observations.
  }
  %
  {
    We compute the temporal evolution of the thermal X-ray emission in kink-unstable coronal loops based on a series of MHD numerical simulations. 
    The numerical setup used consists of a highly twisted loop embedded in a region of uniform and untwisted background coronal magnetic field.
    We let the kink instability develop, compute the evolution of the plasma properties in the loop (density, temperature)  \new{without accounting for mass exchange with the chromosphere.
    We then deduce the X-ray emission properties of the plasma during the whole flaring episode.}
  }
  %
  {
    \new{
    During the initial (linear) phase of the instability plasma heating is mostly adiabatic (due to compression).
    Ohmic diffusion takes over as the instability saturates, leading to strong and impulsive heating (up to more than $20\un{MK}$), to a quick enhancement of X-ray emission and to the hardening of the thermal X-ray spectrum.
    The temperature distribution of the plasma becomes broad, with the emission measure depending strongly on temperature.
    Significant emission measures arise for plasma at temperatures higher than $9\un{MK}$.
    The magnetic flux-rope then relaxes progressively towards a lower energy state as it reconnects with the background flux.
    The loop plasma suffers smaller sporadic heating events but cools down globally by thermal conduction.
    The total thermal X-ray emission slowly fades away during this phase, and the high temperature component of emission measure distribution converges to the power-law distribution $\mathrm{EM}\propto T^{-4.2}$.
    The amount of twist deduced directly from the X-ray emission patterns is considerably lower than the maximum magnetic twist in the simulated flux-ropes.
    }
    }
  %
  {}
  
   \keywords{Sun: corona -- Sun: flares -- Sun: X-rays, gamma rays 
               }

   \maketitle

%
\section{Introduction}
\label{sec:introduction}

\begin{figure*}[]
  \centering
  \includegraphics[width=0.6\linewidth,clip=true,trim=0 100 0  200]{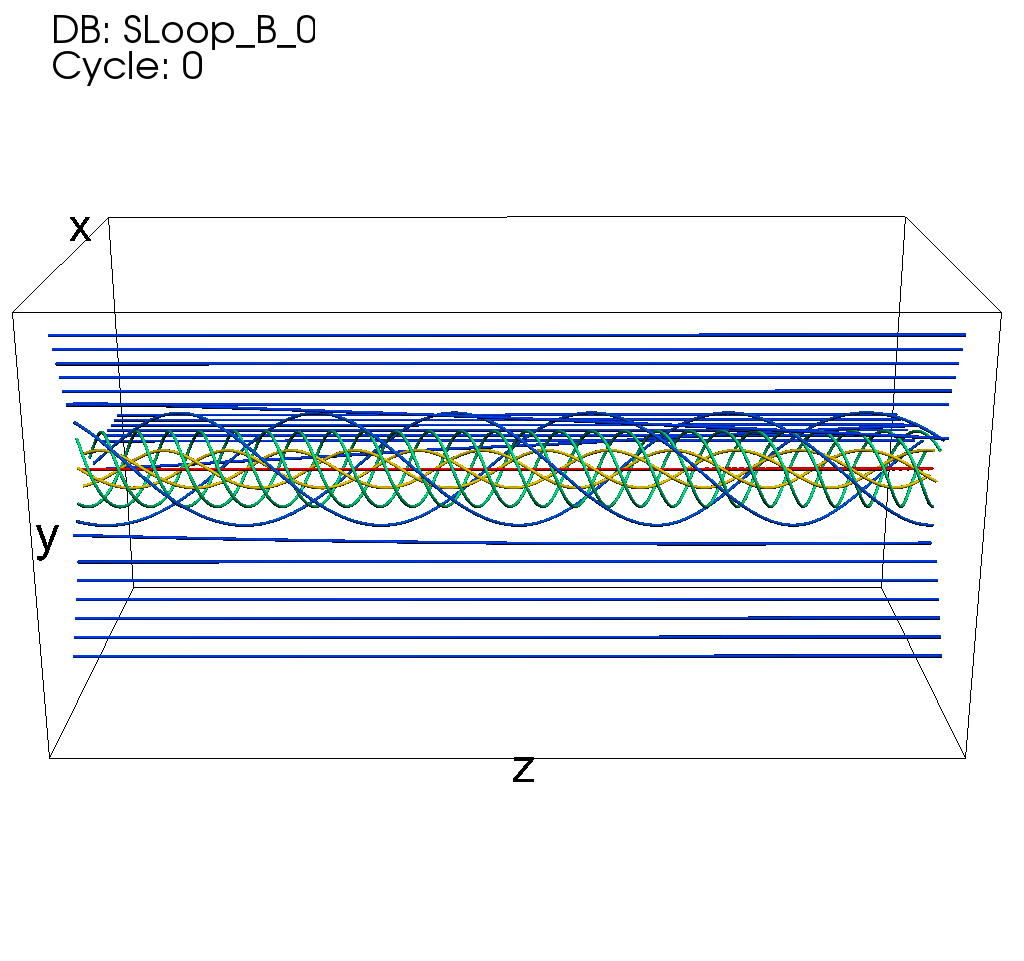} \\ 
  \includegraphics[width=0.75\linewidth,clip=true,trim=0 0 0 13]%
  {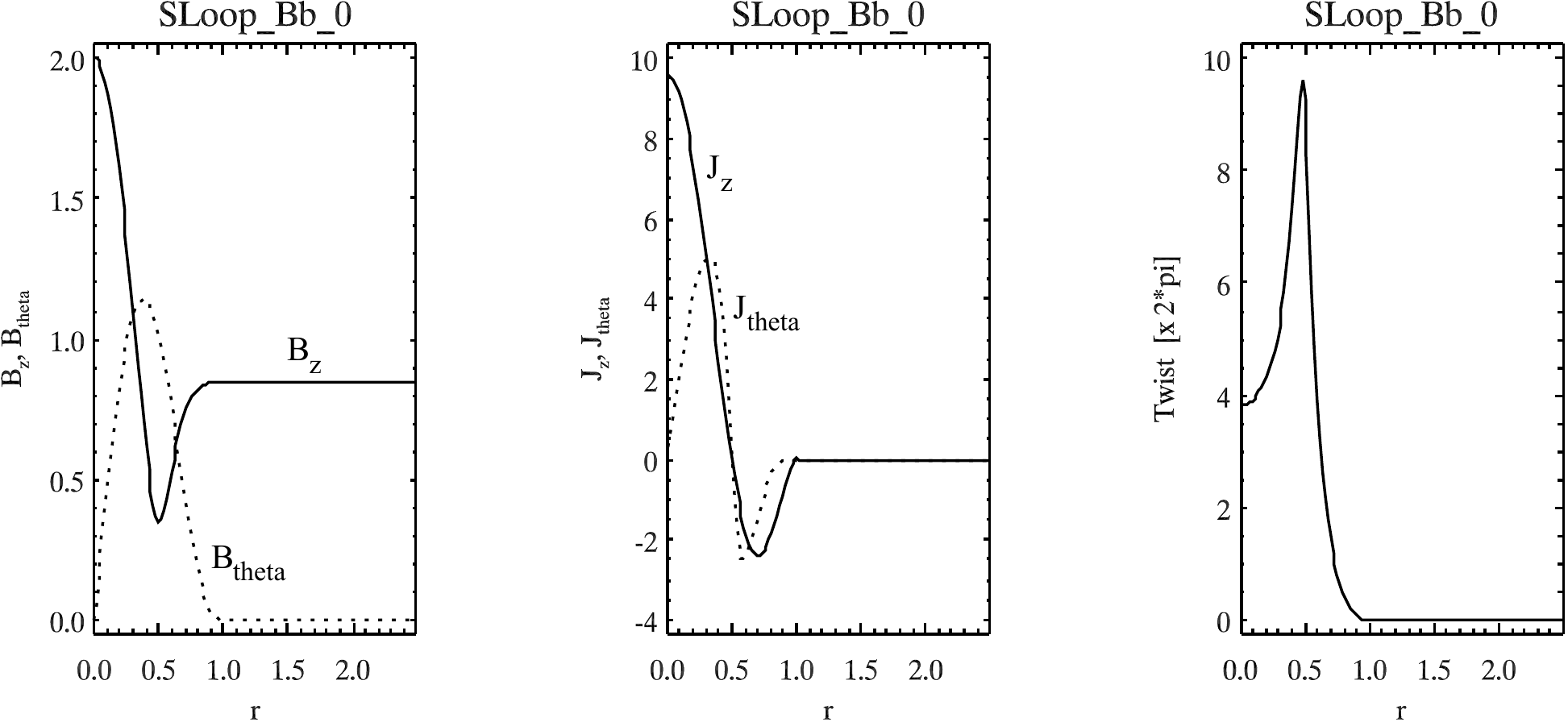}
  \caption{
    The initial conditions for the standard case (magnetic field properties only; see Table \ref{tab:dimensions}).
    The three-dimensional picture on the top show a sample of magnetic field lines (coloured according to the magnetic field strength).
    The plots below show the amplitude of the magnetic field components $B_z$ and $B_\theta$, the current density components $J_z$ and $J_\theta$ (with $\mathbf{J} = \rot{\mathbf{B}} / \mu_0$) and the twist angle $\Phi\left(r\right) = \frac{L_0}{r} \frac{B_{\theta}}{B_z}$ as a function of the radial distance $r$ to the flux-rope's axis.
    All quantities are shown in dimensionless units; these can be converted into physical values (for our standard case) using the rightmost values in Table \ref{tab:dimensions}.
    Continuous lines represent the axial components (along $\hat{\mathbf{e}}_z$) while the dotted lines represent the azimuthal components (along $\hat{\mathbf{e}}_\theta$) of $\mathbf{B}$ and $\mathbf{J}$.
  }
  \label{fig:initial_conditions}
\end{figure*}

Solar flares are energetic phenomena characterised by a quick enhancement of luminosity in a wide spectral range. 
In the soft X-ray domain, in particular, the emitting flux can increase on many orders of magnitudes on time-scales of tens of seconds to minutes \citep{fletcher_observational_2011}.
\new{Solar flares} are usually interpreted as fast releases of magnetic energy stored in the solar corona.
Several energy-storage scenarios are envisioned in the literature.
We shall consider here the scenario in which magnetic energy is stored in twisted magnetic flux-ropes in the corona.
Such magnetic structures are unstable in respect to the kink mode if they are twisted above a certain threshold whose value depends on geometrical properties specific to each individual flux-rope \citep[such as their aspect ratio and transverse pitch angle distribution;][]{bareford_coronal_2013}.
Mechanical perturbations either at their foot-points or at coronal heights may drive them out of their state of equilibrium and trigger the kink instability.
Coronal loops undergoing a kink instability go through an initial linear growth phase until they start reconnecting with the background field \citep{browning_heating_2008}.
They then relax onto a lower energy state (with less twist), hence releasing a fraction of the magnetic free energy stored initially.
This mechanism has been suggested to be at the origin of solar flares of different types (both confined and ejective) and at different spatial scales (from nano-flares to X class flares) \citep[][and references therein]{hood_kink_1979,linton_helical_1996,galsgaard_heating_1997,lionello_nonlinear_1998,shibata_origin_1999,torok_confined_2005,rappazzo_field_2013}.

\new{%
  
  Large flares produce signatures of discrete plasma structures considerably hotter than the coronal background, while smaller and more frequent flares also contribute to the diffuse ``background'' coronal heating.
  The latter kind motivated many studies of coronal loop heating by multiple small amplitude impulsive heating events, such as nano-flares
  \citep[][among many others]{cargill_cooling_2013,bradshaw_influence_2013,west_lifetime_2008,porter_soft_1995,fisher_equation_1990,klimchuk_highly_2008} and turbulent heating \citep{parenti_modeling_2006,buchlin_profiles_2007,verdini_coronal_2012,van_ballegooijen_relationship_2014}.
  The majority of these studies focused on describing in detail the hydrodynamics and heat exchanges occurring in the direction parallel to the magnetic field at the expense of neglecting the transverse gradients and the effects of curvature \citep[see, \emph{e.g}, the review by][]{reale_coronal_2010}.
  This strategy was initially inspired by observations of ultraviolet and X-ray emission structured into multiple thin arched structures connecting regions of opposite polarity at the surface \citep[\emph{e.g}][based on early rocket-launched missions]{vaiana_x-ray_1973}.
}
\new{%
  Improvements to this approach include models of multiple parallel one-dimensional loop strands \citep{reale_trace-derived_2000} and of groups of fine loop strands spreading throughout three-dimensional models of observed active regions \citep{winebarger_verification_2014}, where each strand is treated as a single and independent system.
  These studies have been providing increasingly more sophisticated emission diagnostics, particularly in the EUV range.%
}
\new{%
  However, this \emph{thin thread} approach is less appropriated in the case of strong amplitude oscillations or if the loops undergo magnetic reconnection as is the case in kink-unstable twisted loops leading to larger flares with significant X-ray emission.
}

It is often pointed out that the kink instability scenario requires very large amounts of twist in the flaring coronal loops.
However, observations of flaring coronal loops most often indicate moderate amounts of twist, while highly twisted pre-flare coronal flux-ropes are indeed more rarely observed.
Notable examples exist, nevertheless, such as the highly twisted flux-rope observed by \citet{srivastava_observation_2010}, presumably related to the occurrence of a B5.0 class flare.
The total twist angle of the observed structure is of about $12\pi$ for a loop length of about $80\un{Mm}$ and loop radius $\approx 4\un{Mm}$ (placing it above the Kruskal–Shafranov twist threshold for the kink instability).
%
\new{Additionally, many studies of flux-rope buoyant rise and emergence suggest that highly twisted coronal loops are ubiquitous.}
These twisted magnetic structures are believed to be generated deep inside the Sun's convection zone \new{\citep[see \emph{e.g},][]{nelson_buoyant_2014}}.
Under the right circumstances, they will rise buoyantly across the turbulent convective layers and emerge into the chromosphere and the corona \citep{jouve_three-dimensional_2009,archontis_magnetic_2012,pinto_flux_2013}.
\citet{emonet_physics_1998} have shown that there is a minimum amount of magnetic twist these flux-ropes must have to be able to maintain their coherence during the rise through the convection zone.
Several of these works suggest that this threshold is high enough to be compatible with the kink instability scenario.
This is an important point, as slow helical surface motions alone are unlikely to transmit enough twist to initially untwisted magnetic coronal structures \citep{hood_formulation_1989,grappin_mhd_2008}.
The actual process of transmission of twist up to the corona remains elusive at the present date, though.


\new{In an attempt to link magnetic twist and observations of twisted loops,} \citet{botha_observational_2012} investigated the emission properties in the EUV range of modelled straight coronal flux-ropes undergoing a kink instability by means of numerical MHD simulations.
\citet{gordovskyy_particle_2011} studied the consequences of the reconnection driven by the onset of the kink instability on the acceleration of particles using similar MHD models, and estimated the corresponding hard X-ray signature.

In this paper, we analyse the evolution of the soft X-ray continuum emission in a modelled kink unstable coronal loop.
\new{Our aim is to investigate whether this kind of models is capable of predicting the main properties of soft X-ray emission in solar flares, rather than providing exact reproductions of observations.}
We consider straight twisted magnetic flux ropes which are already kink-unstable.
The determination of the physical mechanisms which lead to the formation of such unstable structures is out of the scope of this manuscript.
\new{We start off from a system already at coronal temperatures and do not address the general problematic of coronal heating by steady or quasi-steady heat sources.
Furthermore, we do not take into account mass transfer between the corona and the chromosphere (hence leaving out the effects of chromospheric evaporation on the density structure of the loops, even during the relaxation phase).}
We chose to use a flux-rope model whose dynamical properties were well studied in the past \citep[\emph{e.g,}][]{hood_coronal_2009,botha_thermal_2011,gordovskyy_magnetic_2011}
and focus on the determination of the properties of thermal continuum emission in the soft X-ray energy range.
We investigate how the spatial distribution of the emitted flux relates to the dynamical and geometrical properties of the simulated loops, how the total continuum emission spectra evolves in time and on how the properties of the emission measures respond to the plasma heating processes occurring in the magnetic loops.

In the remainder of this manuscript, Sect. \ref{sec:methods} describes the methods and model used and Sect. \ref{sec:results} presents the results obtained.
A discussion follows in Sect. \ref{sec:discussion} and a summary of our results is presented in Sect. \ref{sec:summary}.

\section{Methods}
\label{sec:methods}

We study the evolution of kink unstable coronal loops undergoing a flaring episode by means of MHD numerical simulations.
We considered twisted magnetic flux-ropes embedded in a strongly magnetised coronal background. 
The triggering of the kink instability leads to magnetic reconnection  (between the flux-rope and the background magnetic field) and to a burst of plasma heating.
The simulations take into account viscous and ohmic plasma heating, and cooling by thermal conduction.
The properties of the thermal X-ray photon emission are deduced from the temporal evolution of the plasma temperature and density.
Sect. \ref{sec:numerical-code} describes the equations solved and the numerical code used to integrate them.
Sect. \ref{sec:model-parameters} discusses the flux-rope model used, the numerical set-up adopted and our choice of physical parameters.

\subsection{Equations and numerical code}
\label{sec:numerical-code}

We solve the following set of compressible resistive MHD equations
\begin{eqnarray}
  \label{eq:primitives_rho}
  \partial_t \rho & + & \diver{\rho\mathbf{v}} = 0 \\
  \label{eq:primitives_v}
  \partial_t \mathbf{v} & + & \left(\mathbf{v}\cdot\nabla\right)
  \mathbf{v} = -\frac{\nabla p}{\rho} +
  \frac{\mathbf{J}\times\mathbf{B}}{\mu_0\rho} 
  + \frac{\mu}{\rho}\nabla^2\mathbf{v} \\ 
  \label{eq:primitives_e}
  \partial_t e & + & \left(\mathbf{v}\cdot\nabla\right)e = 
  - e \left( \gamma -1 \right) \diver{\mathbf{v}} + \frac{\eta}{\rho} J^2 +  
  \frac{1}{\rho}\diver{\mathbf{q}} \\
  \label{eq:primitives_b}
  \partial_t\mathbf{B} & = & \rot{\left(\mathbf{v}\times\mathbf{B}\right)} - 
  \rot{\left( \eta \mathbf{J} \right)} \ ,
\end{eqnarray}
 where $\rho$ represents the density, $\mathbf{v}$ the flow velocity, $e = \left(\gamma - 1\right) P/\rho$ the specific internal energy of the gas, and $\mathbf{B}$ the magnetic field.
The magnetic resistivity and dynamical viscosity are represented, respectively, as $\eta$ and $\mu$.
The current density is $\mathbf{J} = \rot{\mathbf{B}} / \mu_0$ and $\mu_0$ is the magnetic permeability.

The heat flux $\mathbf{q}$ includes the contributions both from the viscous heat flux $\mathbf{q}_{visc}$ and from the conductive heat flux $\mathbf{q}_c$. 
The former accounts for the dissipation of shear flows, according to
\begin{equation}
  \label{eq:q_visc}
\mathbf{q}_{visc} = \mathbf{v} \cdot \DD \ ,
\end{equation}
where $\DD$ is the isotropic (shear) viscous stress tensor with components 
${\cal D}_{ij}=-2\mu\left[e_{ij}-\frac{1}{3}\left(\nabla\cdot\mathbf{v}\right)\delta_{ij}\right]$,
where $e_{ij}$ is the strain rate tensor.
The latter corresponds to a flux-limited Spitzer-Härm (SH) magnetic field-aligned conductivity.
The SH conductive flux vector is defined as
\begin{equation}
  \label{eq:q_sh}
  \mathbf{q}_{SH} = -\kappa_0 T^{5/2} \left( \hat{\mathbf{b}} \cdot \nabla T \right) \hat{\mathbf{b}}\ ,
\end{equation} 
where $\hat{\mathbf{b}}$ is the unit vector in the direction of the magnetic field and $\kappa_0$ the SH conductivity coefficient \citep{spitzer_transport_1953}.
A correction is applied to this term, so that the conductive heat flux becomes independent of $\nabla T$ for extremely large temperature gradients \citep[see, e.g,][]{orlando_observability_2010,west_lifetime_2008}.
This correction relies on the definition of the saturation flux $q_{sat}$ 
\begin{equation}
  \label{eq:q_sat}
  q_{sat} = \phi \rho c_s^3\ ,
\end{equation}
\new{%
where $\phi$ is an arbitrary coefficient set to $3/2$ in our simulations, $c_s = \sqrt{\gamma T}$ is the sound speed, where $\gamma = 5/3$ is the ratio of specific heats and $T \propto p/\rho$ is the fluid temperature.
  The value of $\phi$ was determined following the analysis by \citet{cowie_evaporation_1977}, and by trying a few different values for this parameter ($\sim 0.1 - 1.5$) and verifying that its variation has a negligible effect on our specific flux-rope setup (we decided to use the upper and more conservative value of the tested parameter-range).
}

The total conductive heat flux is then defined as
\begin{equation}
  \label{eq:q_c}
  \mathbf{q}_c = \frac{q_{sat}}{q_{sat} + q_{SH}} \mathbf{q}_{SH}\ .
\end{equation}
We also consider in some cases an additional right-hand side term for the energy equation (Eq. \ref{eq:primitives_e}) accounting for the radiative losses in the corona.
This term is written as
$- n^2\Lambda\left(T\right)$, 
where $n$ is the numerical density and $\Lambda\left(T\right)$ represents the cooling rate as a function of the temperature for an optically thin plasma.
The value of $\Lambda\left(T\right)$ is obtained from tabulated data calibrated for solar abundances \citep[generated with CLOUDY 90.01,][]{ferland_2013_2013}.
  We do not use this radiative cooling term systematically, as its amplitude is small for the model parameters we chose (see Sect. \ref{sec:model-parameters}), and its inclusion increases significantly the numerical cost of the simulations.
  We nevertheless verified the effects of radiative cooling by running some simulations with this term turned on (see Sect. \ref{sec:thermal-emission}).

The MHD equations (\ref{eq:primitives_rho}) to (\ref{eq:primitives_b}) are solved in dimensionless form using the dimensional scaling factors dependent on the assumed characteristic magnetic field strength $B_0$, the characteristic length-scale $L_0$ and the characteristic density $\rho_0$.
Hence, the characteristic speed is the Alfvén speed $v_0 = B_0 / \sqrt{\left( \mu_0 \rho_0\right)}$, the characteristic time-scale is $t_0 = L_0 / v_0$, the characteristic temperature is given from the equation of state $T_0 = \left(p_0/\rho_0\right)  \left(\mu_H m_p\right)/k_b$ (where $\mu_H$ is the mean molecular mass for a fully ionised hydrogen gas, $m_p$ is the proton mass and $k_b$ is the Boltzmann constant) and the characteristic resistivity is $\eta_0 = L_0 V_0$.

We integrate this set of equations using the numerical code PLUTO \citep[][]{mignone_pluto:_2007}.
Our setup consists of a fixed and uniform cartesian grid with coordinates $x$, $y$ and $z$ such that the magnetic flux-rope is oriented in the $z$-direction.
The foot-points of the magnetic flux-ropes lie on the planes $z=0$ and $z=L_0$.
The system is advanced in time using explicit time-stepping (Hancock scheme associated with a low-diffusion slope limiter) \new{and a \emph{hlld} solver \citep{miyoshi_multi-state_2005}}, except for the diffusive terms (the viscous, resistive and conductive terms).
Spitzer-Härm (SH) thermal conduction, in particular, makes the explicit integration step become prohibitively small. 
For this reason we integrate all the parabolic (\emph{i.e,} diffusive) terms using a Super Time-Stepping implicit scheme (STS), while the other terms follow the usual explicit scheme.
The solenoidal condition ($\nabla\cdot\mathbf{B}=0$) is assured by a hyperbolic divergence cleaning technique \citep{dedner_hyperbolic_2002}.
The boundary conditions are periodic in the $x$ and $y$ directions, and line-tied in the $z$ direction
(\emph{i.e,} the loop's foot-points are line-tied; see Sect. \ref{sec:model-parameters}).
The line-tying condition is applied at the external boundaries (faces) of the outermost numerical cells. 
The velocity, density, pressure and magnetic field are held fixed there.
The diffusive coefficients are null at these boundaries in order to ensure that the magnetic-field remains line-tied.
The velocity gradients are minimised in the first $2$ numerical cells adjacent to the boundaries to ensure numerical stability \citep[in a way similar to that in][]{aulanier_equilibrium_2005}.
A finite conductive heat flux is allowed across the top and bottom boundaries (acting as a proxy to the heat flux from the corona to the chromosphere across the transition region; see Sect. \ref{sec:evol-loop} for a discussion of these effects).

\subsection{Estimating the thermal X-ray emission}
\label{sec:emission}

\new{
We estimate the thermal X-ray emission as a post-processing step based on the spatial distributions of density and temperature obtained from the MHD simulations.
We focus on the continuum emission in the $1 - 25 \un{keV}$ photon energy range (at the low end of the detection range for RHESSI and for the future Solar Orbiter/STIX spectro-imager), and on how its properties evolve in time following reconnection events in the simulated flaring loops (see the discussion in Sect. \ref{sec:caveats}).
}
The continuum thermal X-ray emissivity of a fully ionised hydrogen plasma with uniform number density $n$ and temperature $T$ at a given photon energy $h\nu$ is
\begin{equation}
  \label{eq:emissivity}
  \epsilon\left(h\nu, T\right) = \epsilon_0 n^2 T^{-1/2} g_{ff}\left(h\nu, T\right) \exp{\left( -\frac{h\nu}{k_b T}\right)}\ ,
\end{equation}
where $g_{ff}\left(h\nu, T\right)$ is the Gaunt factor for free-free bremsstrahlung emission and the coefficient $\epsilon_0$ is $6.8\e{-38}$ if the emissivity is to be expressed in $\mathrm{erg\cdot cm^{-3}\cdot s^{-1}\cdot Hz^{-1}}$ \citep{tucker_radiation_1975}.
We use the following piece-wise approximation to the Gaunt factor 
\begin{equation}
  \label{eq:g_ff}
  g_{ff}\left(h\nu, T\right) = 
  \begin{cases}
    1,                                                        & h\nu \lesssim k_bT \\
    \left(\frac{k_bT}{h\nu}\right)^{0.4}, & h\nu > k_b T
  \end{cases}\ .
\end{equation}
The corresponding photon flux density emitted at the photon energy $h\nu$ is defined as
\begin{equation}
  \label{eq:photonflux}
  I\left(h\nu, T\right) = I_0 \frac{\mathrm{EM}}{h\nu \sqrt{k_bT}} g_{ff}\left(h\nu, T\right) \exp{\left( -\frac{h\nu}{k_b T}\right)}\ ,
\end{equation}
where $\mathrm{EM}$ is the emission measure $n^2V$ of a finite volume of plasma (of density $n$ and temperature $T$), and the coefficient $I_0$ is $1.07\e{-42}$ for a photon flux measured at a distance of $1\un{AU}$ and $1.20\e{-41}$ for a photon flux measured at the Solar Orbiter perihelion ($\sim 0.3\un{UA}$), if the photon flux density is expressed in units of $\mathrm{photons\cdot cm^{-2}\cdot s^{-1}\cdot keV^{-1}}$.
The total photon flux over a given spectral band is computed by integrating Eq. (\ref{eq:photonflux}) over the corresponding range of values of $h\nu$.
We compute the photon flux at different photon energies for each individual grid cell (\emph{i.e}, volume element), each one having a one-valued emission measure (note that the density varies in the loop) and temperature.
As the corona is optically thin to X-ray radiation, the total flux emitted is obtained by adding the individual contributions over the whole loop (or over a region of interest).

We estimate the distributions of $\mathrm{EM}\left(T\right)$ in our simulations by computing the total emission measure of the plasma regions whose temperature lies within successive temperature intervals at a given time.
That is, the emission measure is defined as a function of temperature as 
\begin{equation}
  \label{eq:em_t}
  \mathrm{EM}\left(T\right) = \sum_k n^2_k \cdot \delta V_k \ ,
\end{equation}
where the index $k$ runs through all the plasma elements (grid-cells in the simulations) which lie within the temperature interval $\left[T,  T+\delta T\right]$, $n_k$ and $\delta V_k$ are, respectively, the number density and the volume of each element.
In other words, we first compute a temperature histogram with a given temperature bin size $\delta T$.
Then, we verify which grid-cells have a temperature $T$ within each of the bins and we sum over all the corresponding individual $\mathrm{EM}$.
\new{Variations in density in the plasma at a given temperature are therefore accounted for.}

\begin{table}
  \centering
  \begin{tabular}{r c c}
    quantity    & adimensional value &  adopted physical value  \\ 
    \hline\hline
    $L_0$       &               $10$             & $5\e{9}\un{cm}$                \\
    $B_0$      &               $2$                & $2\e{2}\un{G}$                   \\
    $\rho_0$  &               $1$               & $2\e{-14}\un{g\ cm^{-3}}$   \\ 

    \hspace{1em} \\

    $n_0$     &               $1$                  & $1.20\e{10}\un{cm^{-3}}$  \\ 
    $T_0$     &          $5\e{-3}$             & $1.20\e{6}\un{K}$              \\ 
    $\tau_A$&          $5$                     & $1.25\e{1}\un{s}$              \\ 
    $\tau_s$&          $110$                   & $2.75\e{2}\un{s}$              \\ 

    \hspace{.25em} \\

    $\tau_{cond}$&                          & $\approx 1\e{1}\un{s}$                \\
    $\tau_{rad}$  &                          & $\approx 2\e{5}\un{s}$                 \\
    \hline
  \end{tabular}
  \caption{Summary of the model parameters and of the standard choice of physical dimensions (standard case). 
    The first three rows correspond to our choice of independent quantities (loop's length, magnetic field and density), while the following ones are derived from these (numerical density, temperature, Alfvén and sound crossing time-scales).
    The last two rows show the characteristic cooling times (conductive and radiative) at the beginning of the cooling phase.
    See the Sect. \ref{sec:model-parameters} for more details.}
  \label{tab:dimensions}

  \begin{tabular}{l l }
    case          &  new parameters   \\ 
    \hline\hline
    Low-twist  & $\lambda = 2.0 $ \\
    Thin loop  & $r_0 = 0.5$           \\
    Long loop & $L_0 = 20$           \\
    Weak B    & $B_0 = 1$             \\
    Dense loop \hspace{10ex} & $\rho_0 = 4 $  \hspace{4ex} \\
    \hline
  \end{tabular}
  \caption{Summary of the comparative cases and of the parameters which were changed in respect to those in the standard case.
  The new parameters are given in adimensional units, as those in the second column of Table \ref{tab:dimensions}.}
  \label{tab:dimensions_alt}

\end{table}

\subsection{Model, dimensions and parameters}
\label{sec:model-parameters}

We considered here twisted magnetic flux-ropes embedded in a region of uniform background magnetic field aligned with the flux-rope axis direction (the $z$-direction in our setup).
The flux-ropes are straight, and the effects of the large-scale loop's curvature are therefore neglected.
The initial state of the system is force-free and is in hydrostatic equilibrium.
The background medium is characterised by a uniform magnetic field oriented in the $\hat{\mathbf{e}_z}$ direction, a uniform density $\rho$ and a uniform gas pressure $p_0 = \beta B_0^2 / 2\mu_0$, where $\beta$ represents the ratio of gas to magnetic pressures.
We set the parameter $\beta$ to the value $0.01$ in order to correctly represent the dynamics of the magnetically dominated corona.
The twisted magnetic flux-rope is, initially, perfectly cylindrical with its main axis oriented in the $\hat{\mathbf{e}_z}$ direction.
Its characteristic magnetic field is $\mathbf{B} = B_0 \mathbf{\hat{e}_z}$ (at its axis).
Its length $L_0$ matches the numerical domains length and its radius is denoted $r_0$.
The plasma is initially stationary everywhere in the domain ($\mathbf{v} = 0$).
For simplicity, we define the flux-rope magnetic field components in the cylindrical components $B_r$, $B_\theta$ and $B_z$ such that $r$ is the distance to the $z$-aligned flux-rope's axis and $\theta = \tan^{-1}\left( y / x \right)$ is the azimuthal angle.
The radial component $B_r$ is null everywhere.
The components $B_z$, and $B_\theta$ are defined in terms of the twist parameter $\lambda$.
Inside the flux-rope (\emph{i.e,} for $r \leq r_0$)
\begin{eqnarray}
  \label{eq:flux_rope_field_in}
  B_{\theta} &=& B_0 \lambda \frac{r}{r_0} \left( 1 - \frac{r^2}{r_0^2} \right)^3 \\ 
  B_z          &=& B_0 \left[  1 - \frac{\lambda^2}{7}  + \frac{\lambda^2}{7}\left( 1 - \frac{r^2}{r_0^2}\right)^7 - \lambda^2\frac{r^2}{r_0^2} \left( 1 - \frac{r^2}{r_0^2}\right)^6 \right]^{1/2}\ , \nonumber
\end{eqnarray}
and outside (\emph{i.e,} for $r > r_0$)
\begin{eqnarray}
  \label{eq:flux_rope_field_out}
  B_{\theta} &=& 0 \\
  B_z          &=& B_0 \left(  1 - \frac{\lambda^2}{7}  \right)^{1/2}\ , \nonumber
\end{eqnarray}
as in \citet{hood_coronal_2009,botha_thermal_2011,
  gordovskyy_magnetic_2011,gordovskyy_effect_2012}.
The flux-rope's magnetic field matches the background field at $r = r_0$.
The value of the parameter $\lambda$ controls the amount of twist in the flux-rope, rendering it more or less susceptible to the kink instability.
The magnetic field becomes purely axial everywhere in the domain for $\lambda = 0$ (as $B_{\theta} = 0$ and $B_z = B_0$ in that case).
The maximum value of the twist parameter is $\lambda \lesssim 2.438$, ensuring that the square-rooted polynomial in Eq. (\ref{eq:flux_rope_field_in}) is positive.
The flux-rope field is purely toroidal ($B_z=0$) for this limiting value of $\lambda$.
The threshold for the kink instability depends both on the specific transverse twist profile considered and on geometrical parameters such as the flux-rope's aspect ratio \citep{bareford_coronal_2013}.
In our case, and for an aspect ratio $L_0/ r_0 = 10$, the kink instability is prone to develop for $\lambda \gtrsim 1.6$.
Empirically, and given the constraints imposed by diffusive time in our simulations, we verified that cases with $\lambda \leq 1.8$ are impractical to use.
We deliberately chose higher values for the twist parameter ($\lambda=2.0 - 2.4$) in order to guarantee that the kink instability would develop with the least amount of spurious magnetic diffusion.
The system will remain stationary (in its initial state) for an indefinite amount of time unless some form of asymmetry is introduced.
Hence, we introduce a small amplitude seed perturbation in the form of an harmonic velocity noise which is maximal at the centre of the domain and null at the boundaries.
The actual form of the perturbation is unimportant to the outcome of the simulations, as long as its amplitude remains much smaller than the system's characteristic sound and Alfvén speeds.
This mechanical perturbation can be thought of as representing any kind of disturbance in the dynamical corona.


\begin{figure}[]
  \centering
  $t=0\un{s}$ \\
  \includegraphics[width=.9\hsize,clip=true,trim=0 88 0 150]
  {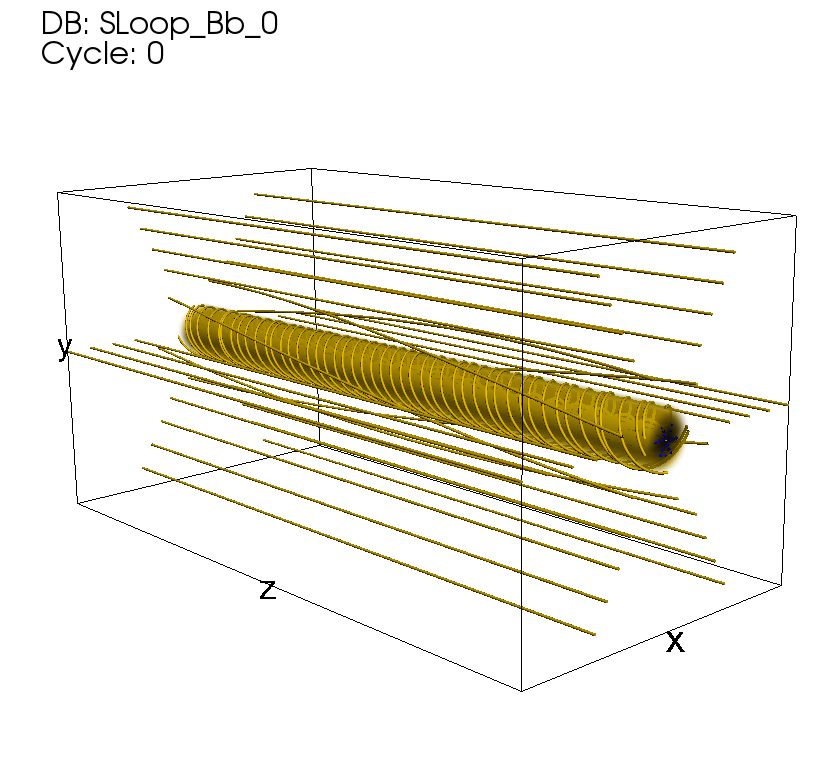}\\
  $t=75\un{s}$ \\
  \includegraphics[width=.9\hsize,clip=true,trim=0 88 0 150]
  {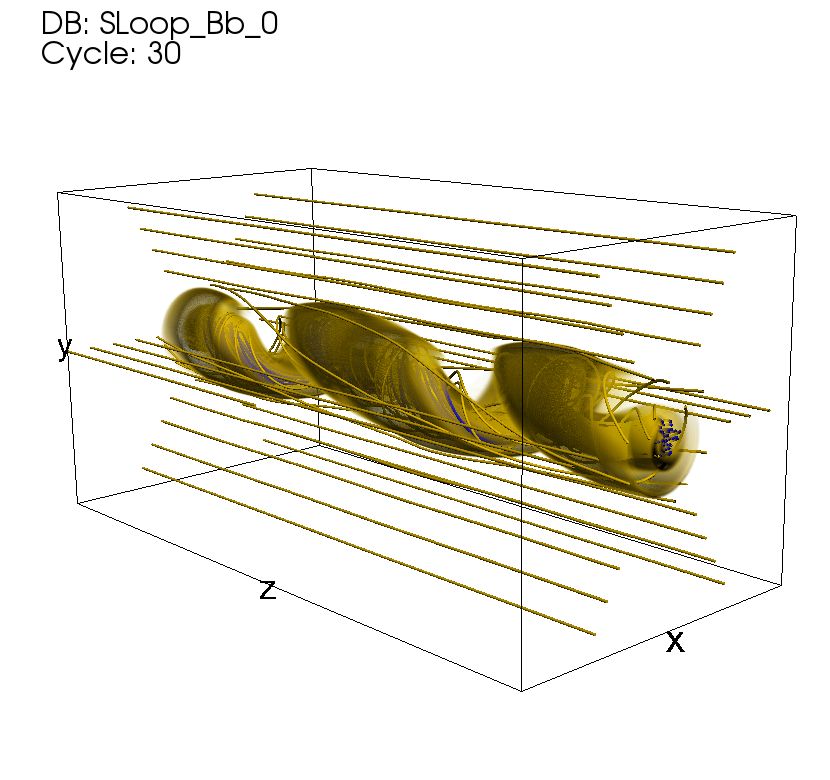}\\
  $t=118\un{s}$ \\
  \includegraphics[width=.9\hsize,clip=true,trim=0 88 0 150]
  {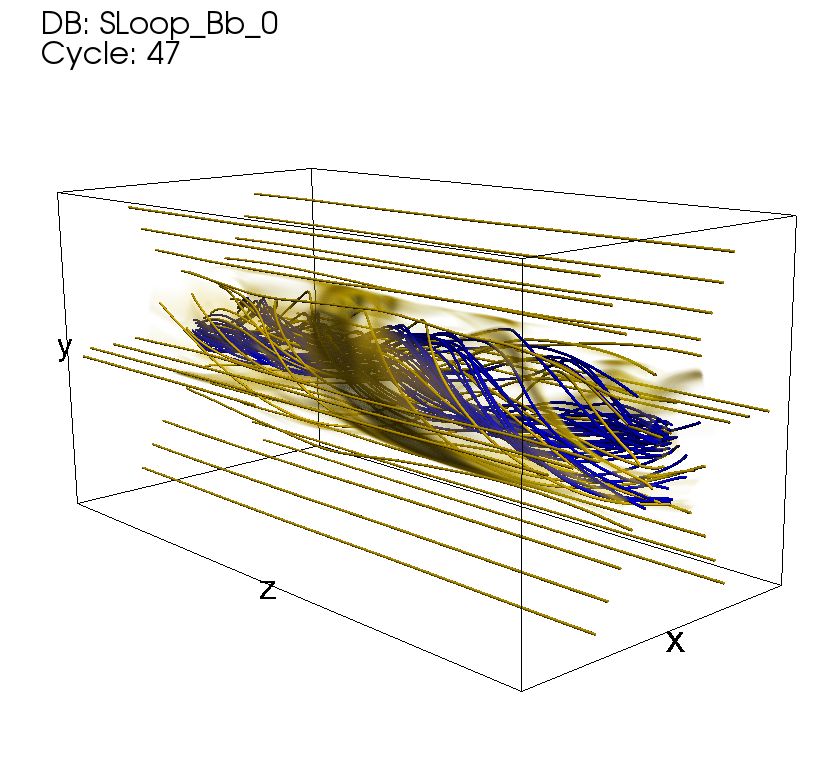}
  \caption{
    Three snapshots showing the temporal evolution of the magnetic field and the current density in the standard case (Table \ref{tab:dimensions}). 
    Blue lines: magnetic field-lines initially placed near the axis of the flux-rope.
    Yellow lines: magnetic field-lines initially crossing the periphery of the flux-rope and the background field. 
    The yellow volumes represent the current density distribution (light/dark yellow corresponding respectively to moderate/strong amplitudes).
    The inner (blue) magnetic field-lines are concentrated well within the current-carrying region (hence hidden in the first two panels) before the reconnection event takes place.
    The instants represented correspond to the initial state ($t=0\un{s}$),  to the peak in magnetic energy release rate ($t=75\un{s}$)  and to the relaxation phase ($t=118\un{s}$). 
    }
    \label{fig:Blines}
\end{figure}

\begin{figure}[!h]
  \centering
  \includegraphics[width=.9\hsize,clip=true,trim=0 0 0 8]{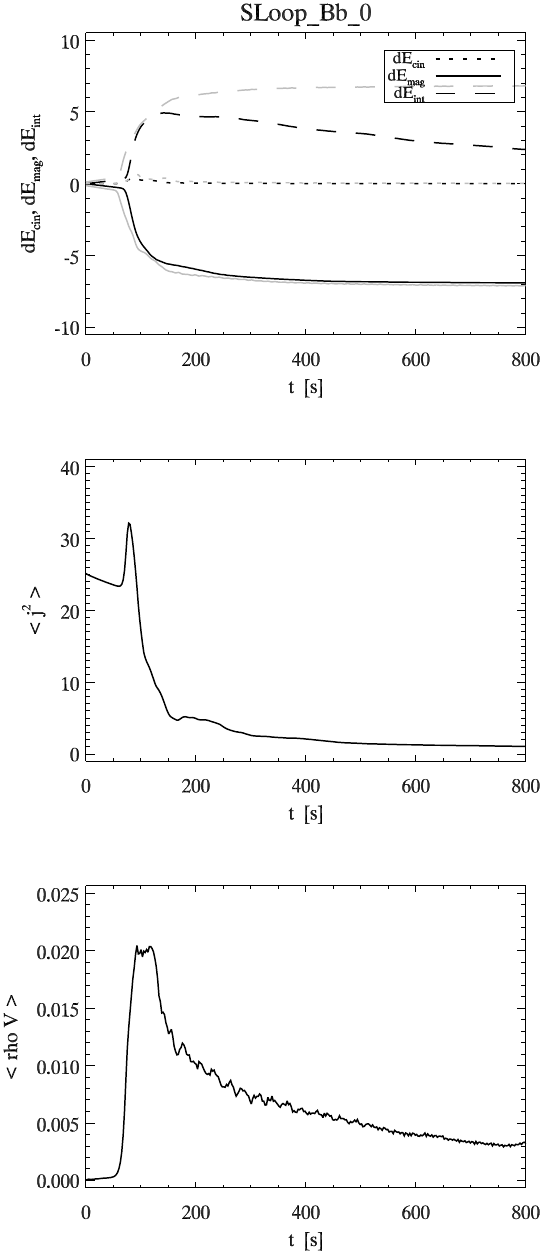}  
  \caption{
    Temporal evolution of the variations in total kinetic, magnetic and internal energies ($\Delta E_{cin}$, $\Delta E_{mag}$, $\Delta E_{int}$), average current density squared $\langle j^2 \rangle$, and average linear momentum $\langle \rho v \rangle$ in the standard case (in dimensionless units).
    The grey lines in the top panel represent the same quantities for a case without thermal conduction, for comparison.
    The total kinetic energy is always small in respect to the magnetic and internal energies.
    The thermal conductive flux then starts growing fast as the plasma quickly heats up locally (due to the ohmic dissipation),  and is responsible for the decay in internal energy during the relaxation phase (note that in our setup the conductive flux can transport heat outwards through the loop's footpoints).}
    \label{fig:energies}
\end{figure}

Figure \ref{fig:initial_conditions} shows a three-dimensional rendering of the magnetic field at the initial state of one of our simulations. 
The blue and green lines represent magnetic field-lines connected, respectively, to the twisted flux-rope and the background field.
The figure to the right shows the distribution of the twist angle $\Phi\left(r\right) = \frac{L_0}{r} \frac{B_{\theta}}{B_z}$ in a plane perpendicular to the flux-rope's axis.
Note that the twist profile $\Phi\left(r\right)$ is controlled by the parameter $\lambda$, and that the effects of varying the latter can translate into qualitatively different twist distributions.

The rectangular numerical grid has a length $\Delta l_z = L_0 = 10$ and a width $\Delta l_x = \Delta l_y = 5$ in normalised units\new{, and the grid dimension is $256^3$, with the grid-cells thinner in the transverse $x$ and $y$ directions than on the longitudinal $z$ direction.
Other resolutions were tested, such as $256\times 256\times 512$ and $512\times 512\times 1024$ (i.e, different grid-cell sizes and aspect ratios) to verify numerical stability.}
The flux-rope radius is, in the standard case, $r_0 = 1$.
The characteristic magnetic field strength $B_0$ equals $2$, the characteristic density equals unity, and the characteristic temperature is $T_0 = 5\e{-3}$.
The flux-rope's Alfvén and sound longitudinal crossing time-scales therefore are $\tau_A = 5$ and $\tau_s \approx 110$.
In order to scale the simulations to coronal values, we set $B_0^c = 200\un{G}$, $L_0^c = 5\e{9}\un{cm}$ and $\rho_0^c = 2\e{-14}\un{g\cdot cm^{-3}}$.
As a consequence, the coronal temperature is $T_0^c = 1.2\un{MK}$ \new{(a typical coronal loop temperature)} and the Alfvén and sound crossing times are, respectively $\tau_A^c = 12.5\un{s}$ and $\tau_s^c = 275\un{s}$.
\new{The typical size of the grid-cells then is $\sim 100\un{km}$ in the transverse directions ($x$ and $y$) and $\sim 200\un{km}$ in the longitudinal direction ($z$).}
The flux-rope's viscous and resistive time-scales $\tau_\eta = a^2/\eta$ and $\tau_\mu = \rho a^2/\mu$ are $\sim 500 \tau_A$.
\new{%
  The magnetic resistivity is uniform, with a value $2\e{14}\un{cm^2 s^{-1}}$, which more than ensures the stability of the numerical scheme, with magnetic Reynold's numbers never larger than $\sim 1$ at the grid-scale \citep[for comparison, this value is close to those reported by, \emph{e.g},][]{bingert_intermittent_2011}.
As a consequence, the bulk magnetic diffusion at large scales is non-negligible, and low-twist scenarios become harder to calculate than high twist cases.
}
This set of parameters ensures that the twisted flux-ropes are kink unstable, and that they are strongly and quickly heated during the first phase of the evolution of the instability (see Sect. \ref{sec:results}), after which they go through a cooling phase.
%
\new{We expect plasma cooling to be dominated by thermal conduction rather than by radiation for our typical loop parameters during the dynamical time-scales we will be considering ($\sim 10^{2} - 10^{3}\un{s}$; the estimated conductive to radiative cooling time-scales being of about $10^{-5} - 10^{-4}$ during that period).
Hence, we will not account for the latter on the majority of the cases studied.
We verified \emph{a posteriori} that this assumption was correct (see Sect. \ref{sec:thermal-emission} and Fig. \ref{fig:lightcurves_cooling}).}
It should be noted, nevertheless, that the cooling time-scales depend strongly on the choice of model parameters. 
For example, substantially denser and colder loops could reach higher values for the ratio $\tau_{cond}/\tau_{rad}$, or even switch from conductively-cooled to radiatively-cooled regimes during the course of the relaxation phase \new{\citep[see, \emph{e.g},][]{cargill_implications_1994,cargill_cooling_1995,klimchuk_highly_2008}}.
We will not consider such cases here.
Table \ref{tab:dimensions} shows a summary of our standard choice of dimensional scaling parameters, which we will hereafter refer to as standard case.
Variations to the standard case will be referred to according to the names in Table \ref{tab:dimensions_alt}.

\section{Results}
\label{sec:results}

We describe hereafter the different stages of the temporal evolution of the simulated kink unstable loops.
The main geometric features and the global dynamical behaviour of the system are, as expected, in good agreement with previous studies \citep[\emph{e.g,}][]{hood_coronal_2009,botha_thermal_2011,gordovskyy_magnetic_2011}.
Our main contribution to this body of research lies on the study of the properties of thermal X-ray emission on such systems.
Section \ref{sec:evol-loop} describes the temporal evolution of the magnetic field and currents during the flaring episode, and describes the overall energy balance. 
Section \ref{sec:thermal-emission} describes in detail the X-ray emission properties.
Section \ref{sec:dem} describes the development of a multi-temperature plasma in the flaring loops.
These results are discussed in respect to X-ray observations of flaring loops in Section \ref{sec:discussion}.

\subsection{Dynamical evolution}
\label{sec:evol-loop}

\begin{figure*}[!ht]
  \centering

  


  $t = 75\un{s}$ \\
  \includegraphics[width=0.51\textwidth,clip=true,trim=0 100 0 150]%
  {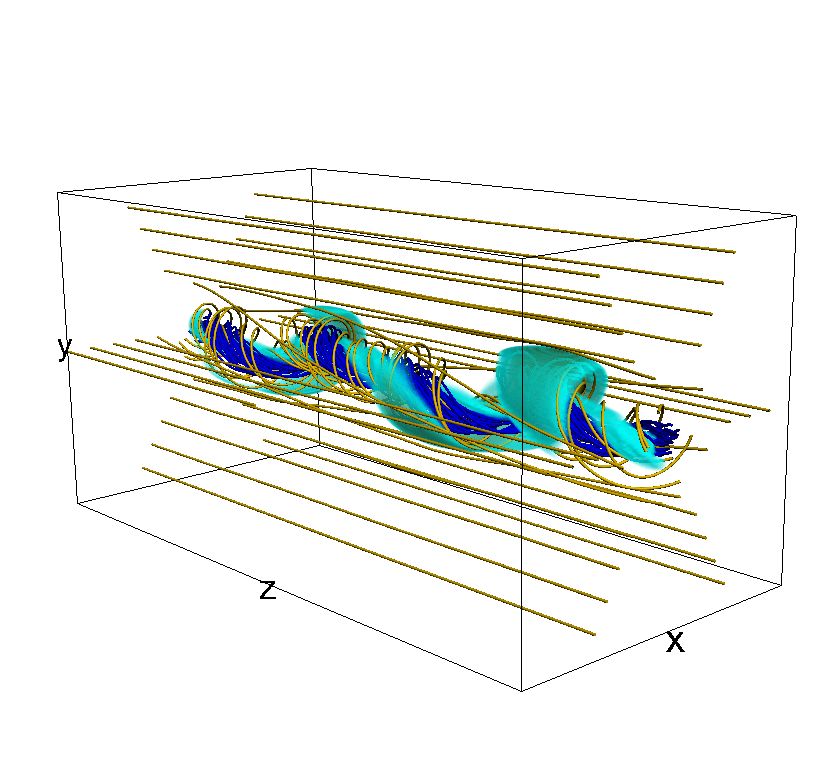} \hspace{0.02\textwidth}
  \includegraphics[width=0.42\textwidth,clip=true,trim=0      0 0   20]%
  {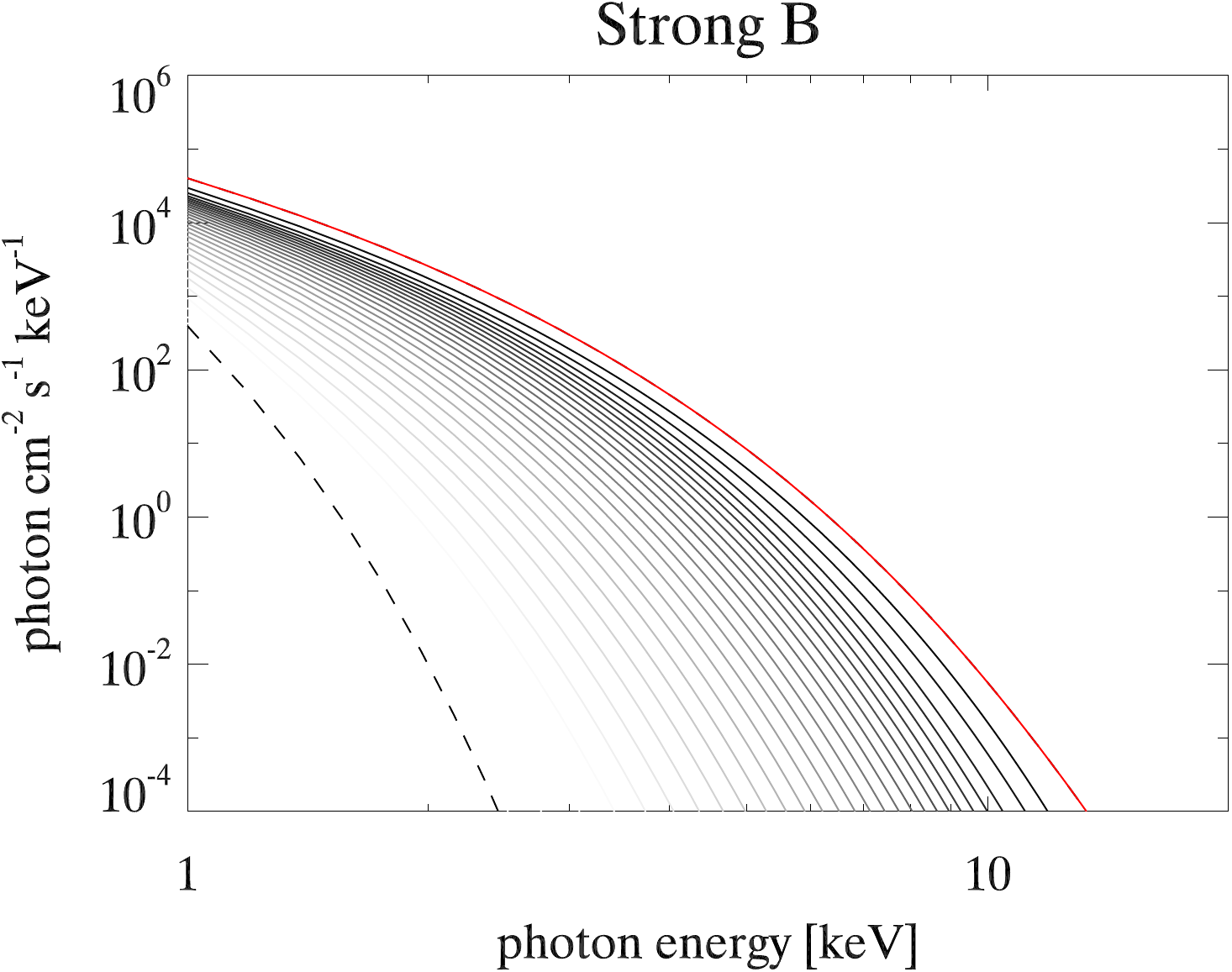} \\ \vspace{0.035\textwidth}
  
  $t = 90\un{s}$ \\
  \includegraphics[width=0.51\textwidth,clip=true,trim=0 100 0 150]%
  {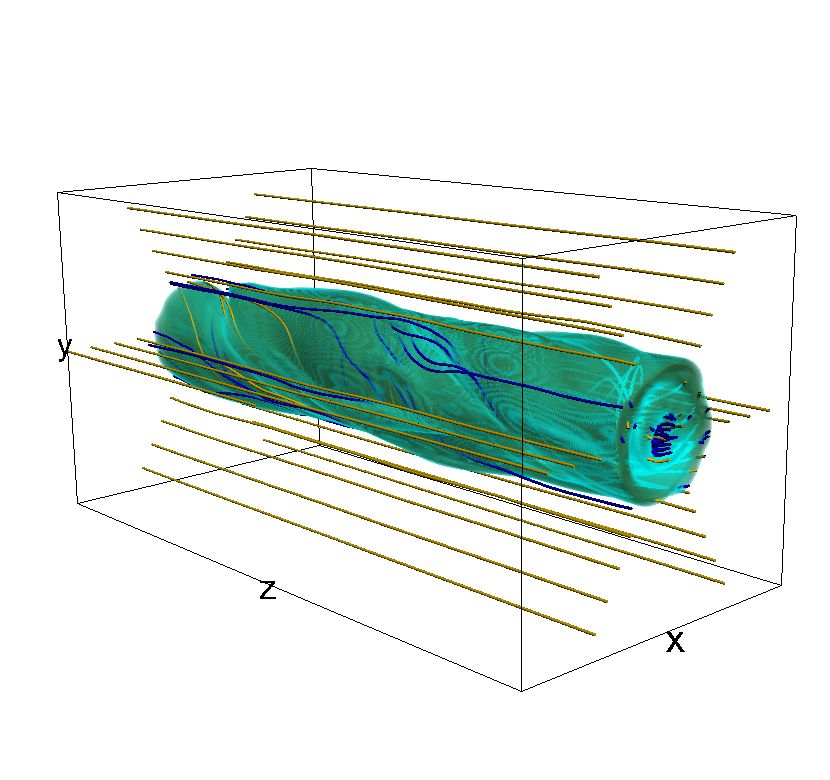} \hspace{0.02\textwidth}
  \includegraphics[width=0.42\textwidth,clip=true,trim=0      0 0   20]%
  {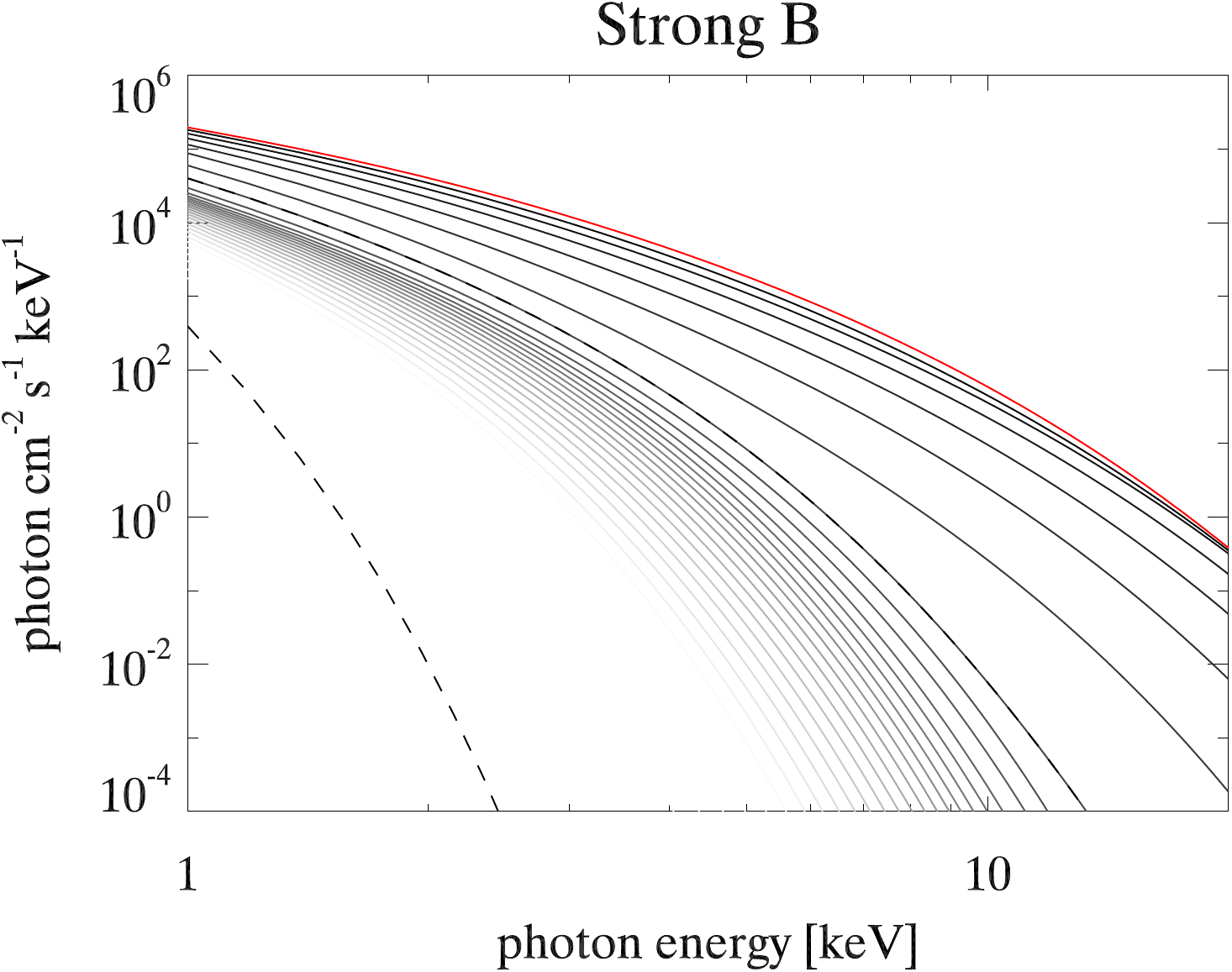} \\ \vspace{0.035\textwidth}

  $t = 475\un{s}$ \\
  \includegraphics[width=0.51\textwidth,clip=true,trim=0 100 0 150]%
  {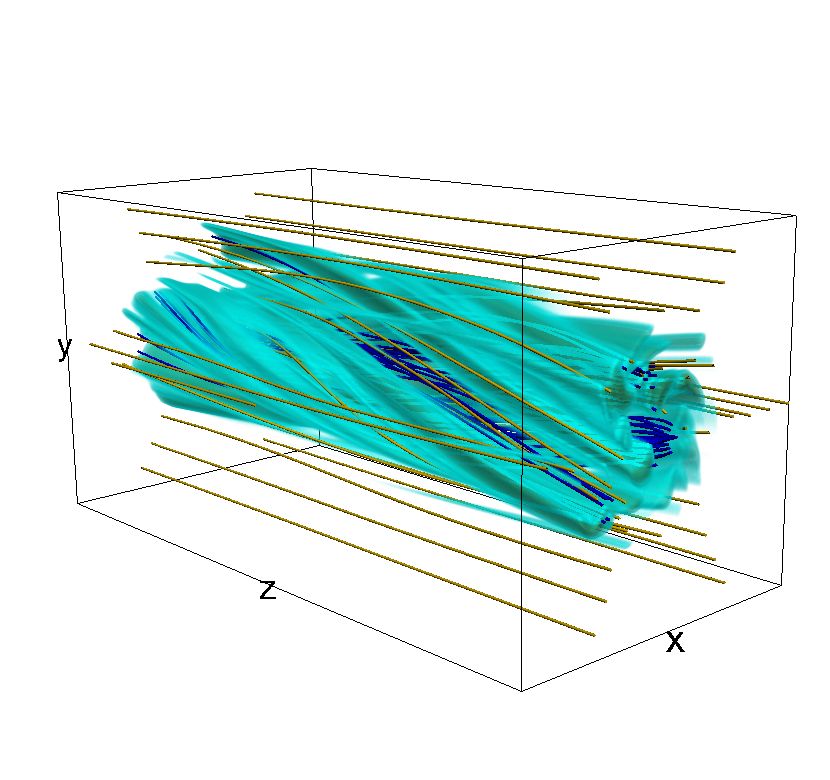} \hspace{0.02\textwidth}
  \includegraphics[width=0.42\textwidth,clip=true,trim=0      0 0   20]%
  {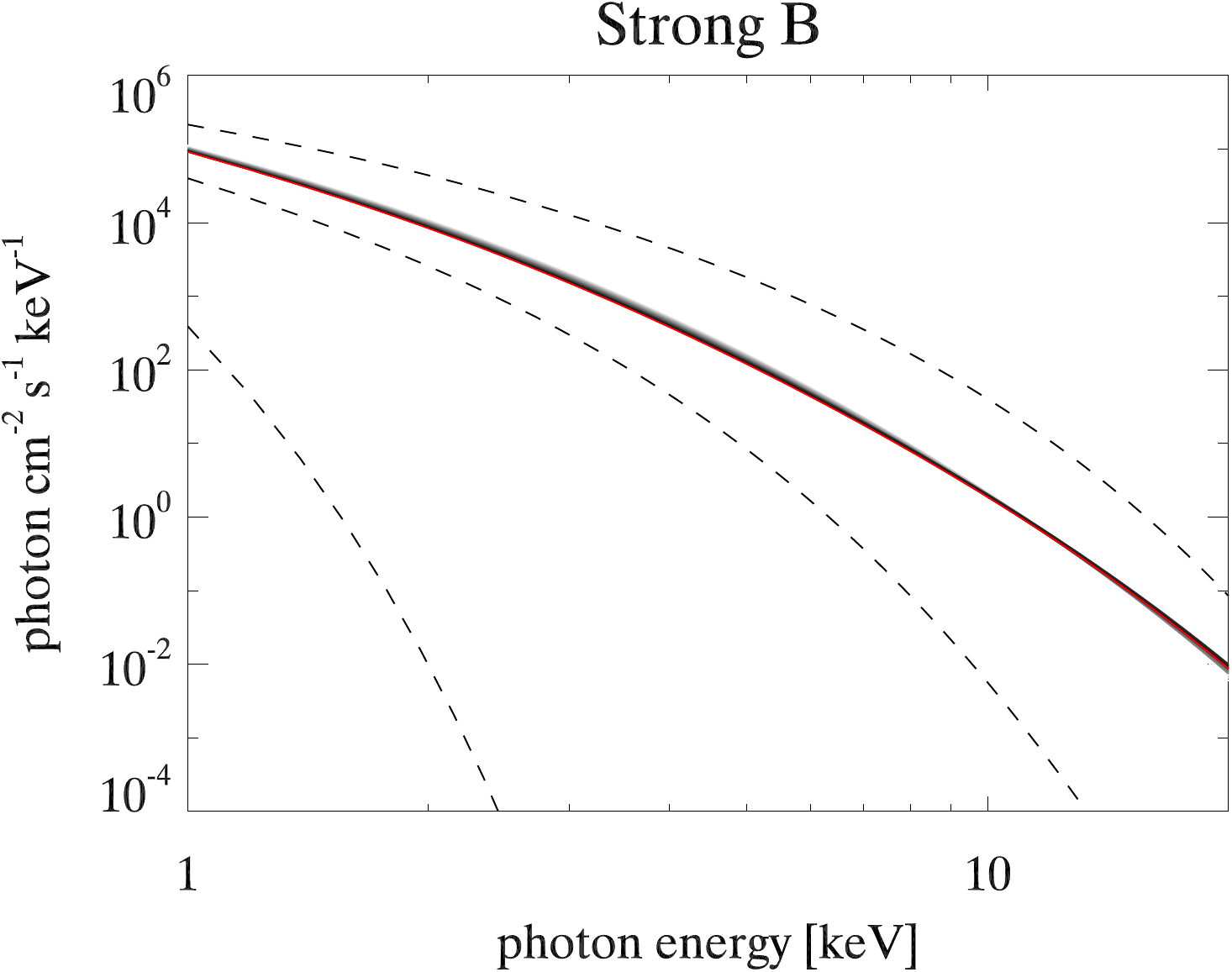} 

  \caption{
    \new{Temporal evolution of the magnetic field, of the emissivity at $10\un{keV}$ (as defined in Eq. \ref{eq:emissivity}) and of the total emission spectrum (Eq. \ref{eq:photonflux}) in the standard case.}
    The instants represented correspond, from top to bottom, to the linear phase ($t = 75\un{s}$), the saturation phase ($t = 90\un{s}$) and the relaxation phase ($t = 475\un{s}$).
    The left column shows three-dimensional renderings of the magnetic field (blue and yellow lines, as in Fig. \ref{fig:Blines}) and of emissivity (green volumes) at these instants.
    The  right column shows the corresponding emission spectra between $1$ and $20\un{keV}$.
    The red lines show the spectra at the same instants as the figures to the left, and the light to dark grey show spectra at some preceding instants (with $2.5\un{s}$ of time-delay between each line) hence giving an idea of the quickness of the evolution of the spectra.
    The black dashed lines show the initial ($t=0\un{s}$), end of linear phase ($t\approx 75\un{s}$) and peak spectra ($t\approx 90\un{s}$).
    }
    \label{fig:emission_and_spectra}
\end{figure*}

The temporal evolution of the kink unstable twisted flux-rope is divided in three distinct phases, which we will name hereafter the linear phase, the saturation phase and the relaxation phase.
Figure \ref{fig:Blines} shows a few snapshots illustrative of these phases.
The yellow and blue lines represent magnetic field-lines rooted at the top and bottom boundaries, and the yellow volumes represent the current density distribution.
The twisted flux-rope is initially at rest, and the kink instability is triggered after an initial perturbation breaks its perfect cylindrical symmetry (perturbing its magnetic tension balance).
From then on (and as long as the linear phase of the instability lasts), the flux-rope kinks about its axis and expands outwards.
The plasma is heated by compression ahead of the boundaries between the expanding regions and the background medium.
Helical-shaped and thin current sheets form and grow at these interfaces.
At a certain point, the flux-rope magnetic field starts reconnecting with the background field, and the linear instability (exponential growth) saturates.
The system's magnetic geometry is quickly reconfigured, the peripheral current sheets start fragmenting and decaying in amplitude, and strong and localised heating occurs there.
The saturation time-scale depends directly on the values assumed for the diffusive coefficients  and weakly on the amplitude of the initial perturbation.
In particular, lower magnetic resistivities will allow the plasma compression to proceed for a longer period of time and lead to stronger peak currents.
In our numerical setup, the saturation time-scale is of the order of $4 - 5$ Alfvén crossing times.
From then on, the global magnetic field will slowly converge to a state with lower twist, closer to a potential field configuration.
During the saturation and relaxation phases, the current density looses its initially smooth and cylindrically symmetric distribution (see the plots in Fig. \ref{fig:initial_conditions} and the first image in Fig. \ref{fig:Blines}) and assumes a more intermittent spatial distribution, until it eventually fades away.

Figure \ref{fig:energies} shows the absolute variations of total kinetic, magnetic and internal energies in the system as a function of time, as well as the temporal evolution of the average current density and momentum.
The total kinetic, magnetic and internal energies are defined, respectively, as $E_{cin} = \frac{1}{2}\int_V \rho v^2 dV$, $E_{mag} = \frac{1}{2\mu_0}\int_V  B^2 dV$ and $E_{int} = \int_V \rho e dV$.
The overplotted grey lines show the same quantities but for a model without thermal conductivity (Eqs. \ref{eq:q_sh} to \ref{eq:q_c}).
The initial excess of magnetic free energy is predominantly transferred into thermal energy, while only a small fraction is converted into kinetic energy.
The plasma flow velocities remained small at all times (below $0.1 c_s$), despite the initial impulsive acceleration and the low plasma $\beta$.
The variations of total internal energy $\Delta E_{int}$ and total magnetic energy $\Delta E_{mag}$ are almost perfectly reciprocal during the linear phase (and especially so in the non-conductive case).
During the saturation phase, the strong and localised increases in plasma temperature make the thermal conduction very efficient. 
As a result, the maximum $\Delta E_{int}$ attained is slightly reduced in respect to the non-conductive case.
From then on, the total internal energy will slowly decay and approach its initial value.
This happens because thermal conduction is allowed to let heat flow through the loop's foot-points in our setup.
If this was not the case, the magnetic loops would reach higher maximum temperatures and would almost not cool down after the saturation phase, keeping $\Delta E_{int}$ at a stable level.
%
The current density peaks at about $t=110\un{s}$, which corresponds roughly to the instant represented in the second panel in Fig. \ref{fig:Blines}, just before the magnetic reconnection event starts.
The current density quickly decays from that moment on (as the magnetic field is reconfigured and relaxes).
The plasma, initially at rest, is accelerated (essentially outwards) during the linear phase of the instability.
After reconnection is triggered, some longitudinal acceleration appears for a short period of time in some places of the flux-rope.
This is at the origin of the second peak in the momentum curve (third panel in Fig. \ref{fig:energies}).
These large scale bulk flows are attenuated during the subsequent relaxation phase, but the smaller scale flows persist, composing a mildly turbulent medium.
Longitudinally propagating low-amplitude oscillations triggered during the initial burst survive for a long period of the relaxation phase (at least up to $t=1000\un{s}$).

\subsection{Thermal X-ray emission}
\label{sec:thermal-emission}

\new{We focus now on the properties of the thermal bremsstrahlung X-ray emission deduced from our simulations}.

\begin{figure}

   \includegraphics[width=.75\hsize,clip=true,trim=0 0 0 10]%
   {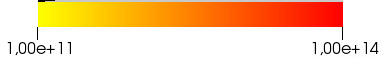} \\ 

   \centering

   $t=65\un{s}$ \\
   \includegraphics[width=.95\hsize,clip=true,trim=0 250 0 250]%
   {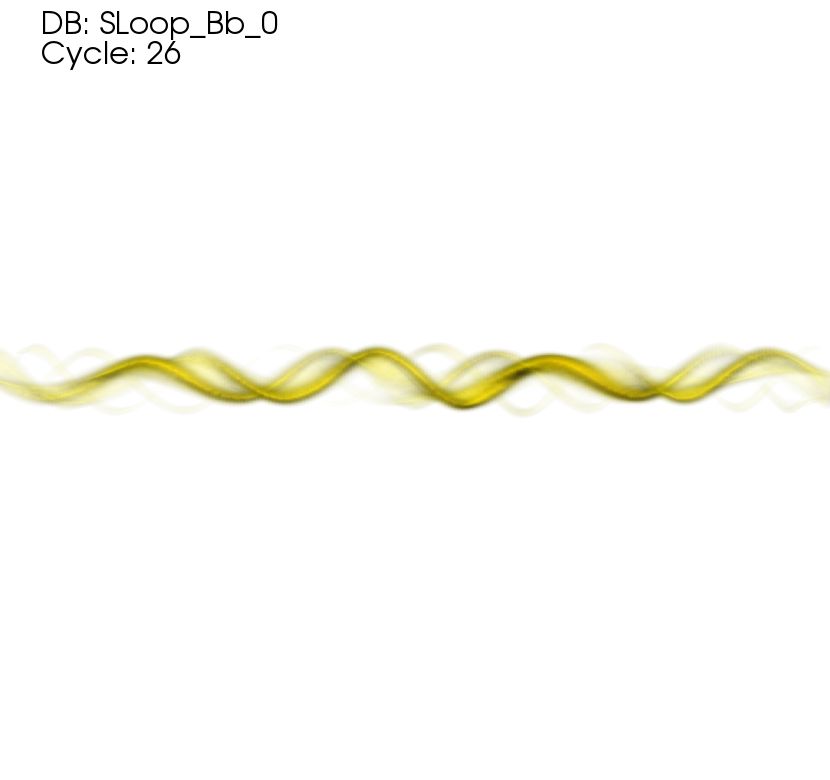} \\

   $t=75\un{s}$ \\
   \includegraphics[width=.95\hsize,clip=true,trim=0 250 0 250]%
   {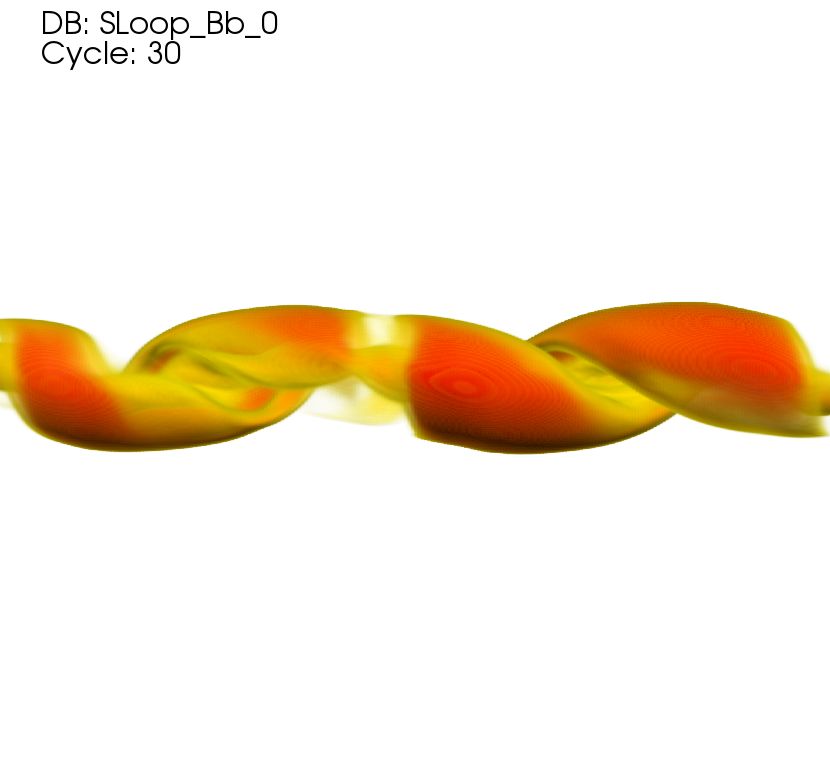} \\

   $t=245\un{s}$ \\
   \includegraphics[width=.95\hsize,clip=true,trim=0 250 0 250]%
   {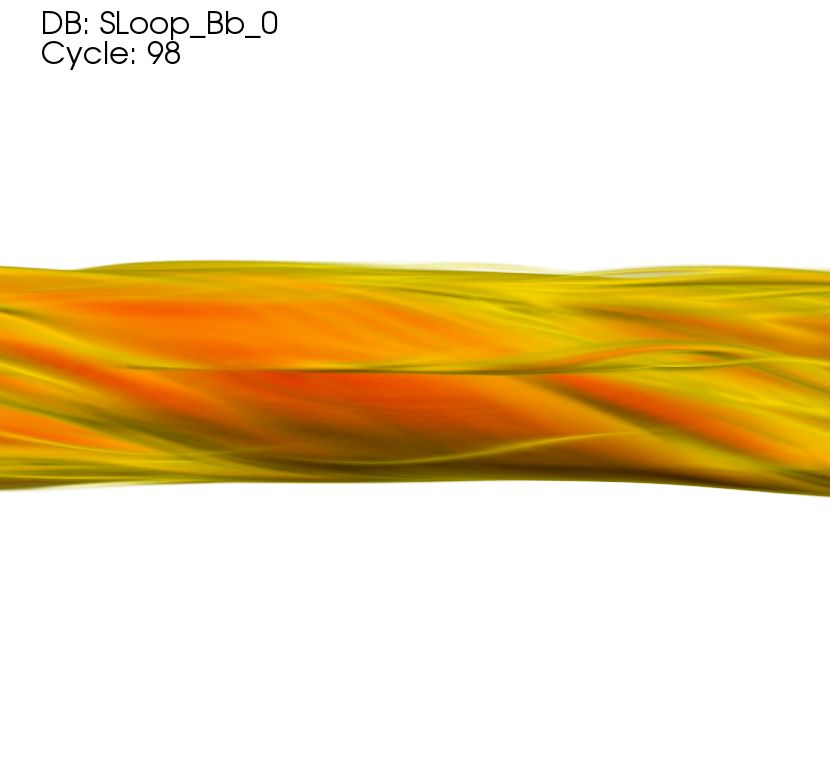} \\

   $t=475\un{s}$ \\
   \includegraphics[width=.95\hsize,clip=true,trim=0  200 0 250]%
   {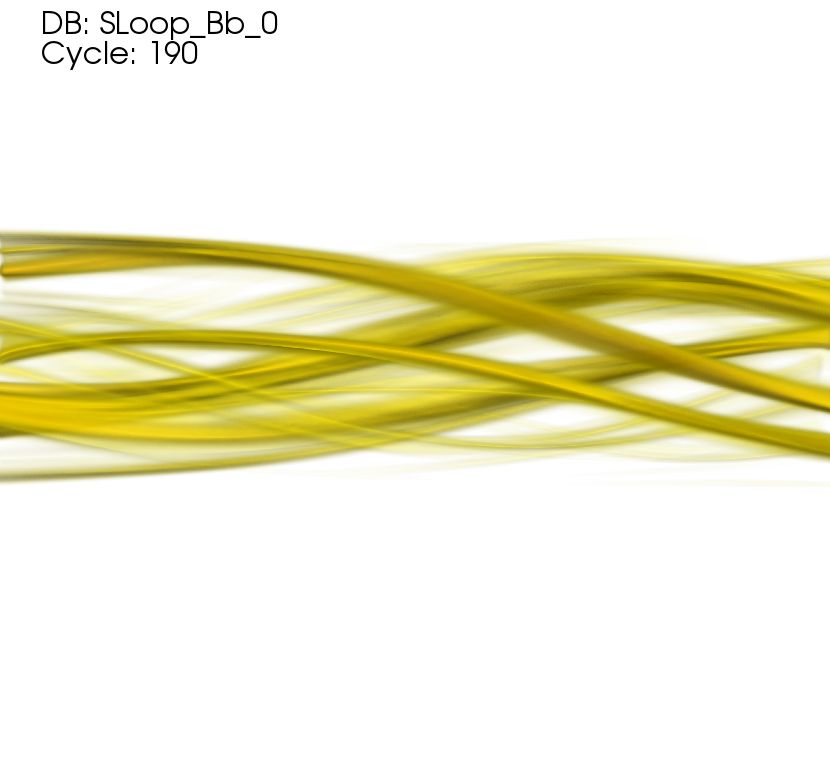} \\
   \includegraphics[width=.95\hsize]%
   {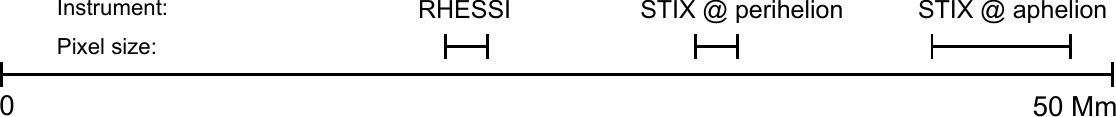} \\
   
   \caption{
     Detail of the continuum emission at $5\un{keV}$ at different instants for the standard case.
     The orange/red colour-table represents the emissivity\new{, as defined in Eq. (\ref{eq:emissivity})}.
     Dark red represents $\epsilon = 10^{13}\un{erg\ s^{-1}\ cm^{-3}\ Hz^{-1}}$, a factor $10$ stronger than yellow.
     The instants represented correspond, in order, to the final moments of the linear phase ($t=65\un{s}$), to the saturation phase / peak of emission ($t=75\un{s}$), to the early relaxation phase ($t=245\un{s}$) and to the later relaxation phase ($t=475\un{s}$).
     The scale on the bottom shows the corresponding pixel size for RHESSI and STIX both at the aphelion ($\sim 1\un{AU}$) and perihelion ($\sim 0.3\un{AU}$) of the spacecraft's orbit (see movie online).
   }
   \label{fig:loop_closeup}

\end{figure}

\begin{figure}
   \centering
   $t=65\un{s}$ \\
   \includegraphics[width=.95\hsize,clip=true,trim=0 250 0 250]%
  {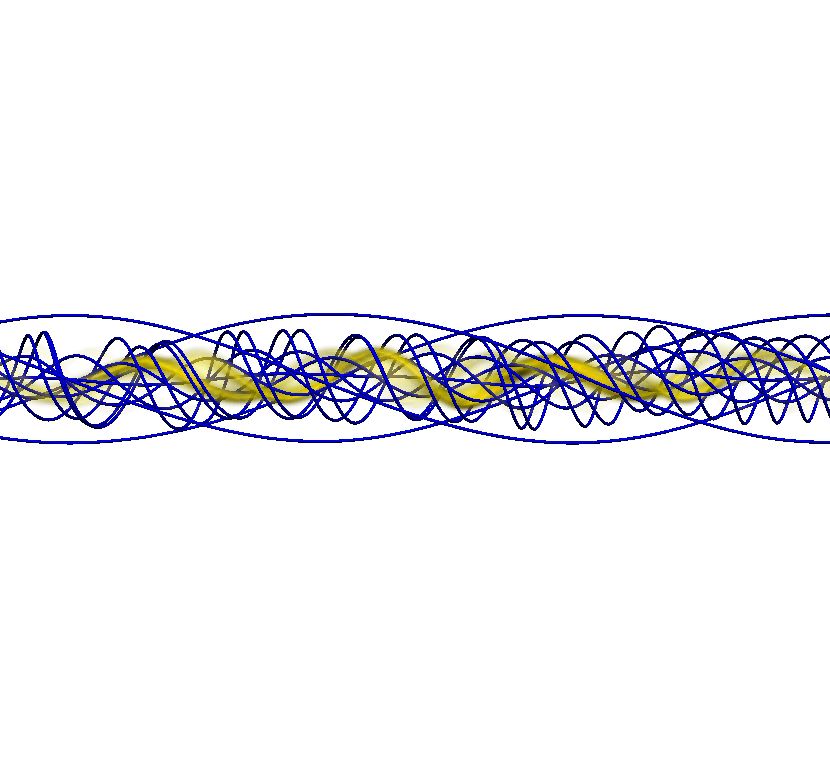} \\

   \caption{
     The first instant in Fig. \ref{fig:loop_closeup} plotted together with some magnetic field-lines (also for the standard case, and with the same colour-table).
     The thermal X-ray emission pattern highlights only parts of the flux-rope with low-twist during the initial phases of the flare.
     Later on, the emission pattern ends up filling all the flux-rope volume but the flux-rope will have lost much of its twist by then.
     On the overall of the flare evolution, the most highly twisted magnetic field-lines are very rarely visible.
   }
   \label{fig:loop_closeup_blines}





\end{figure}

Figure \ref{fig:emission_and_spectra} shows a sequence of three snapshots of the magnetic field and of three-dimensional renderings of the emissivity at $10\un{keV}$ \new{(see Eq. \ref{eq:emissivity})}, accompanied by the photon spectra at $1\un{AU}$ \new{(see Eq. \ref{eq:photonflux})} at the same instants by the total loop volume.
The instants represented are $t = 75\un{s}$ (end of the linear phase), $t = 90\un{s}$ (during the saturation phase and peak of emission), $t = 475\un{s}$ (during the relaxation phase).
The blue and yellow lines represent, respectively, magnetic field lines initially within the twisted flux-rope and the background field (as in Fig. \ref{fig:Blines}).
The green volumes represent the regions of the plasma emitting strongly at $10\un{keV}$.
The red lines on the plots to the right of the figure show the total photon spectra at $1\un{AU}$ at the same instants as the figures to the left, and the light to dark grey lines show spectra at some preceding instants (hence giving an idea of the quickness of the evolution of the spectra).
The time interval between consecutive grey lines is $2.5\un{s}$.
The black dashed lines show the spectra at the initial state ($t=0\un{s}$), at the end of the linear phase ($t\approx 75\un{s}$), and the peak spectra ($t\approx 90\un{s}$).

The first signs of thermal emission appear close to the axis of the flux-rope in small discontinuous patches heated by compression.
These very quickly extend along the corresponding magnetic field-lines, hence forming filamentary and helical emission pattern which highlights the writhe (large-scale ``twist'') of the flux-rope.
A strong helical current sheet then starts forming around the flux-rope (in the zones more strongly compressed against the external medium; see Fig. \ref{fig:Blines}).
The ohmic heating grow quickly there, and the emission concentrated in these outermost layers overcome the latter (\emph{cf.} the second panel in Fig. \ref{fig:loop_closeup}).
The emission is enhanced rapidly as the kink instability proceeds, filling the adjacent zones and forming a compact and continuous emitting structure (see the second row in Fig. \ref{fig:emission_and_spectra}, left side, at $t=90\un{s}$).
During the initial phases (for $t\leq 65\un{s}$), the photon flux spectrum is increased at a steady pace (first row in Fig. \ref{fig:emission_and_spectra}, plot on the right side), mostly as a result of the moderate compressional heating taking place at the emitting regions. 

When the saturation phase is reached, the magnetic field is quickly reconfigured by reconnecting with the external field (see the second row in Fig. \ref{fig:emission_and_spectra}, $t=90\un{s}$).
A strong ohmic heating burst accompanies the reconnection process.
The total photon flux increases very sharply (note the larger gaps between consecutive lines in the plot to the right), and the photon spectrum increases dramatically at high energies. 
This spectral hardening occurs as the loop's plasma transitions from a state with nearly uniform temperature to a state with a broad temperature distribution (with an extended upper tail, reaching as high as $\sim 35 \un{MK}$; \emph{cf.} Sect. \ref{sec:dem} and Fig. \ref{fig:inverse_fit}).

During the relaxation phase, the emission starts decaying slowly, while becoming again more fragmented and concentrated in field-aligned filaments (last row in Fig. \ref{fig:emission_and_spectra}).
The plasma cools down globally (see Fig. \ref{fig:energies}), but maintains its multi-thermal character for a very long period of time (see Sect. \ref{sec:dem}).
The photon spectrum hence decays as a whole approaching the initial one, but without becoming as soft as initially for the whole duration of the simulations.

The fine structure of the emission is shown in more detail in Fig. \ref{fig:loop_closeup}.
The 4 panels represent volume renderings of the emissivity at $5\un{keV}$.
The colour table covers a factor $10^3$ in emissivity, from light yellow to dark red (see the inset colour-scale at the top of the figure).
The scale at the bottom indicates the length of the loop ($50\un{Mm}$) as well as the corresponding pixel sizes for RHESSI and STIX (both at aphelion and perihelion of the orbit planned for the Solar Orbiter spacecraft).
The instants represented illustrate the final moments of the linear phase ($t=65\un{s}$), the saturation phase ($t=75\un{s}$, when the current density is maximal), the beginning of the relaxation phase ($t=245\un{s}$) and a later moment of the relaxation phase ($t=475\un{s}$).
The first traces of thermal emission appear oriented along a few (and only a few) magnetic field-lines.
This happens because the thermal conductivity transports heat efficiently along the magnetic field (and not across), and because the plasma heating sources are initially very discontinuous in space.
\new{Note that the initial mechanical perturbation breaks the initial cylindrical symmetry, and that the plasma becomes reasonably turbulent from early on.
Hence, reconnection is not forced in a perfectly symmetrical way.
Emission displays nonetheless a rather symmetrical large-scale pattern.}.

For a short period of time, the bulk of the thermal emission effectively traces a few magnetic field-lines in the flux-rope.
The emission pattern then displays a clear helical pattern, with a perceived twist of about $\approx 3$ turns (or a twist angle of $6\pi$) at this instant.
Note that this value is much lower than the initial magnetic twist angle in this region ($8-18\pi$; see Fig. \ref{fig:initial_conditions}).
This difference occurs for two reasons.
The first is that the loop has already lost an important fraction of its twist when the plasma becomes hot enough to produce distinguishable emission patterns.
The second reason is that the corresponding field-lines are well within the twisted flux-rope, and hence have lower pitch angles (and lower total twist) than the more external ones (see the twist radial profiles in Fig. \ref{fig:initial_conditions}).
This effect is seen more clearly in Figure \ref{fig:loop_closeup_blines}, where a sample of magnetic field-lines is plotted together with the emission pattern at the same instant.
The bulk of the emission is indeed concentrated close to the kinking flux-rope axis, and is surrounded by more strongly twisted field-lines (for which there are no traces of emission at this moment).
This emission pattern is only visible for a brief period of time ($10 - 15 \un{s}$), though.
The second panel in Fig. \ref{fig:loop_closeup} shows the moment when ohmic dissipation in the helical current sheet formed around the kinking flux-rope becomes important (compare with the second panel in Fig. \ref{fig:Blines}).
Note that the emission peak does not occur at the same time as the peak in ohmic heating rate (proportional to $\left| j^2 \right|$; see Fig. \ref{fig:energies}), but a little while after (about $20\un{s}$ after).
The third panel shows the beginning of the relaxation phase, when the emission becomes almost cylindrically symmetric.
The last panel represents the later relaxation phases.
The emission traces again a more threaded pattern, qualitatively similar to those in coronal loops observed in the EUV range.
The apparent radius of the loop (the radius of the emitting plasma) varies as a function of time in our simulations \citep[\emph{cf.}][]{jeffrey_temporal_2013}.
The width of the emitting region is initially smaller than that of the actual magnetic flux-rope, but increases quickly as the instability proceeds. 
It then stabilises during the relaxation phase as the emission fades away.
%
The typical widths of the filamentary emission patterns described above are below the maximum spatial resolution obtained by RHESSI or by the future STIX instrument (Solar Orbiter).
The overall helical structure would probably be unnoticed by these instruments.
Later on and for most of the flaring episode, the emission pattern should be visible only as a cylindrically symmetric structure (with no apparent traces of helicity).

\begin{figure}[t]

  \centering
  \includegraphics[width=\hsize,clip=true,trim=0 0 0 25]{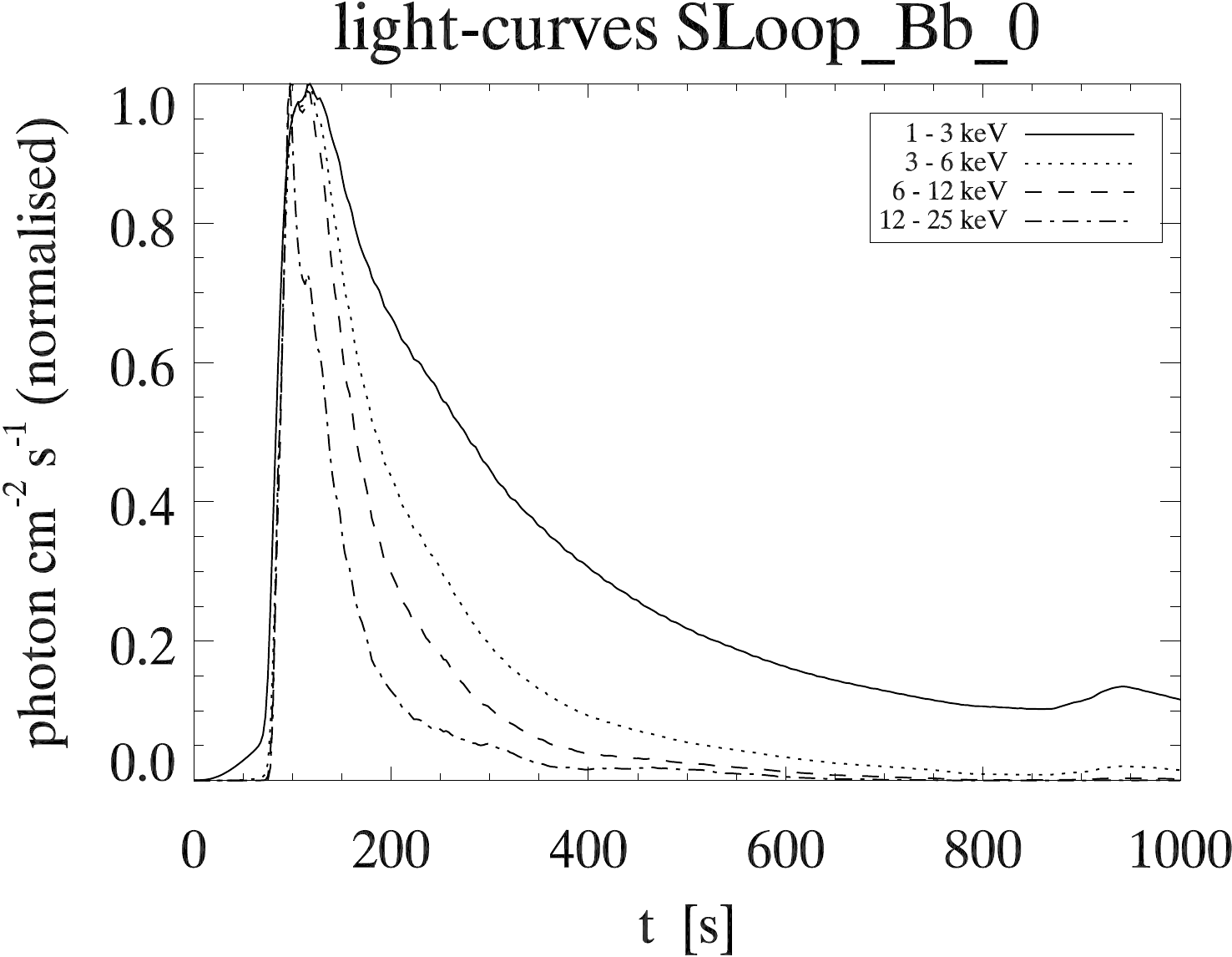}
  \caption{
    Light-curves of X-ray thermal emission at different energy bands for the standard case.
    The curves are all normalised to their peak value.
    The inset key shows which energy band corresponds to each line in the plot.
    Higher energy bands decay faster, lower energy bands decay more slowly.
    The maximum time lag between different peaks is of about $10\un{s}$ (the lowest energy bands peaking earlier).
  }
  \label{fig:lightcurves}
\end{figure}

Figure \ref{fig:lightcurves} shows a series of light-curves computed for different photon energy bands.
The light-curves were computed by integrating the spectra in Fig. \ref{fig:emission_and_spectra} over each energy band at all instants.
Emission from the whole numerical domain was taken into account.
The light-curves are normalised to their maximum value in order to facilitate the comparison (the peak values differ by more than two orders of magnitude between the lowest and the highest energy bands).
The thermal X-ray emission peaks during the impulsive phase and decays asymptotically during the relaxation phase.
The light-curves peak almost synchronously.
The maximum time lag between different peaks is just slightly higher than $15\un{s}$, with the highest energy band ($12 - 25 \un{keV}$) preceding all the others, and with the lowest energy band ($1 - 3 \un{keV}$) being last.
However, the lower energy light-curves are observed to start growing earlier and more progressively than the higher energy ones. %
More importantly, the decay time-scale is longer for the lower energy bands and shorter for the higher energy bands.
This is consistent with the fact that thermal conduction damps the high temperature peaks efficiently, as the conductive flux is proportional to $T^{5/2}\nabla T$ (see eq. \ref{eq:q_sh}).

\begin{figure}[t]
  \centering
  \includegraphics[width=.98\hsize]{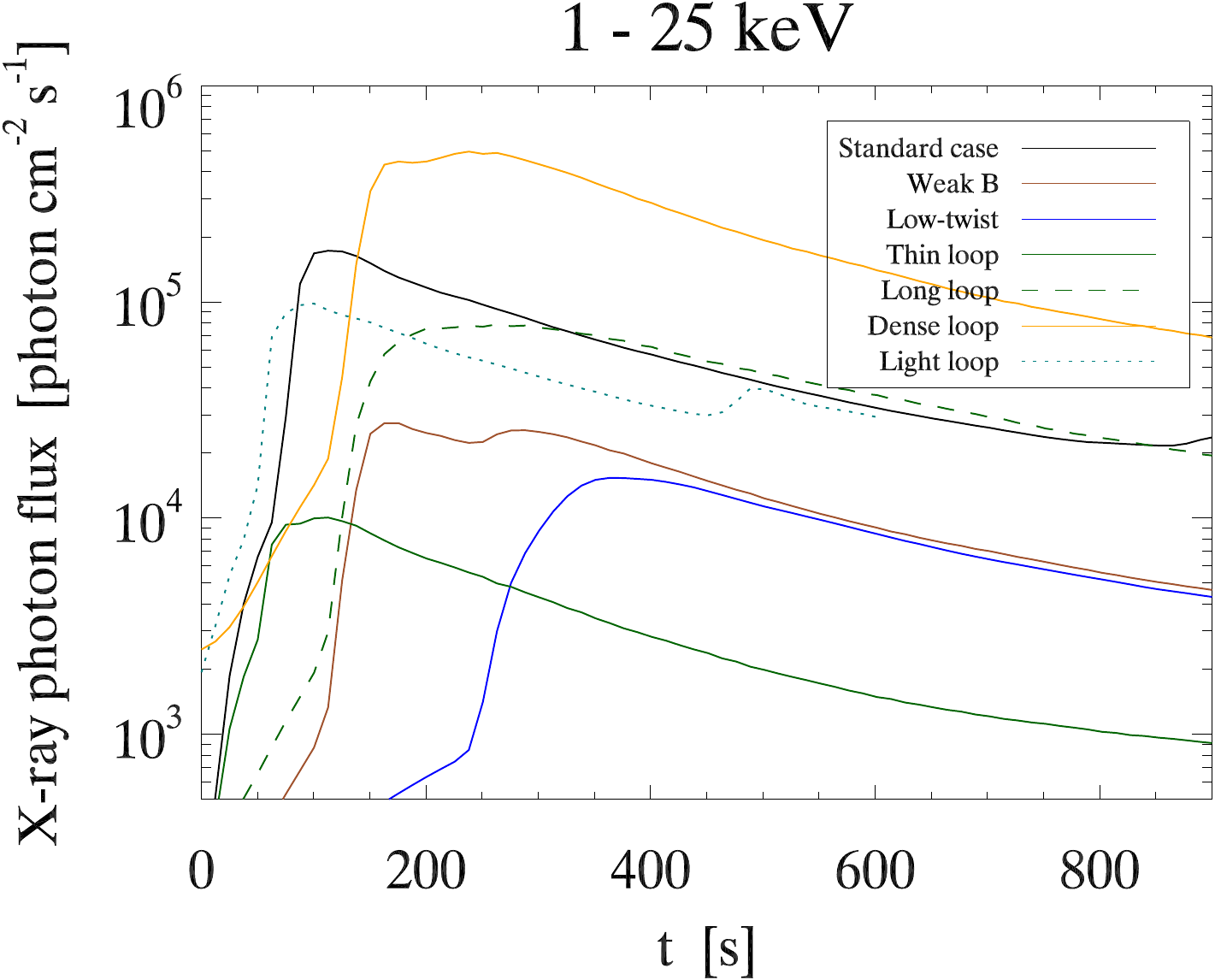} \\ \medskip
  \includegraphics[width=\hsize]{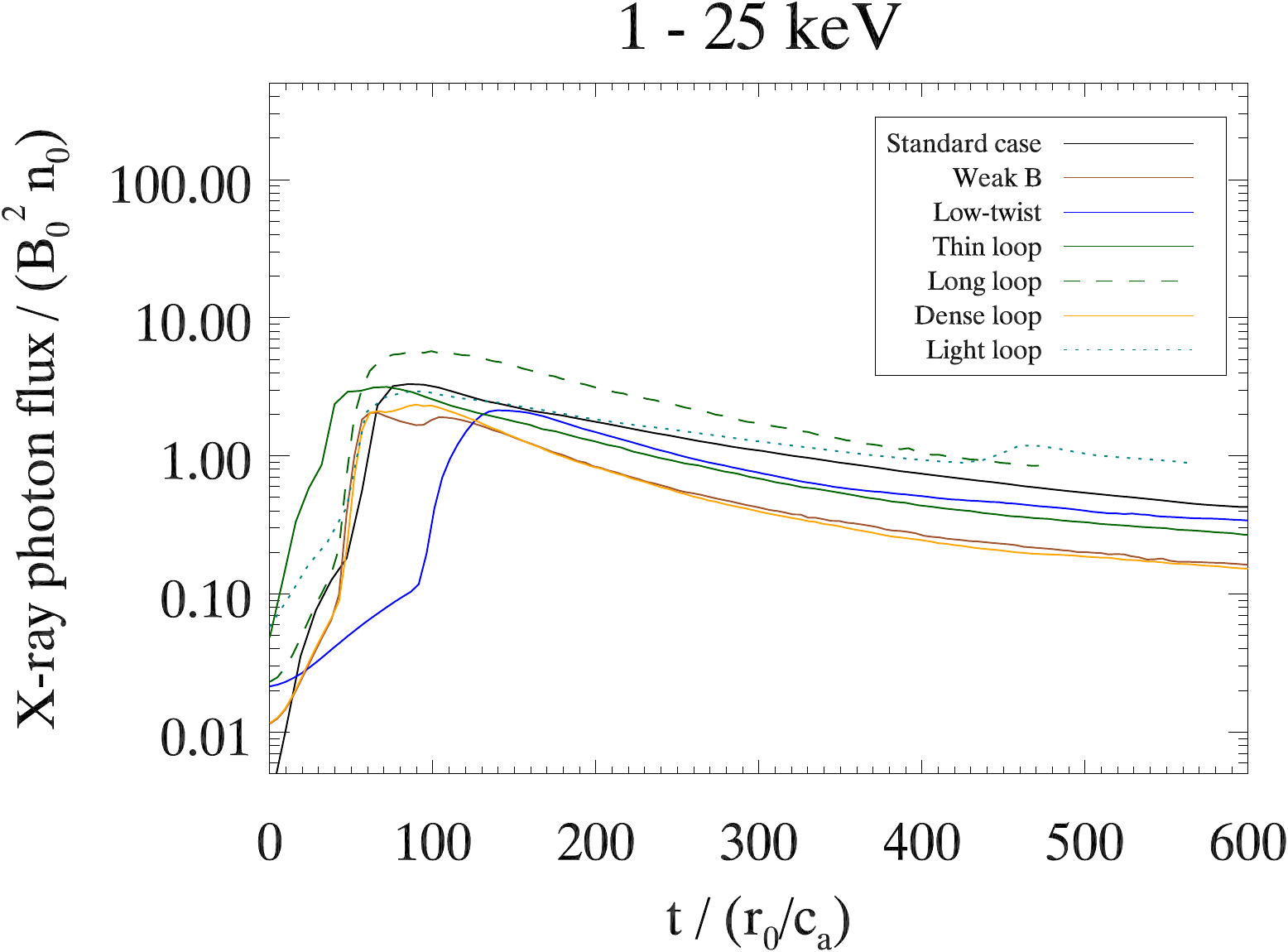}
  \caption{
    Top panel: Light-curves for different cases in the $1-25\un{keV}$ range.
    The cases represented are the standard case, a non-conductive case, a low-twist case, a case with a flux-rope twice as thin, a case with a flux-rope twice as long, a strong $\mathbf{B}$-field case (twice as strong), and a case with strong $\mathbf{B}$-field and a denser flux-rope such that the plasma beta and characteristic Alfvén speed are maintained (see the inset legend).
    Bottom panel: The same light-curves but with rescaled time and photon flux.
    The maximum of emission is approximately proportional to $B_0^2  n_0$ (product of the initial magnetic energy and of the initial density) for each case.
    The time-scale to attain the emission peak is proportional to $r_0/v_0$ (ratio of flux-rope radius to characteristic Alfvén speed).
  }
  \label{fig:lightcurves_comparative}

\end{figure}

\begin{figure}[t]
  \centering
  \includegraphics[width=.95\hsize]{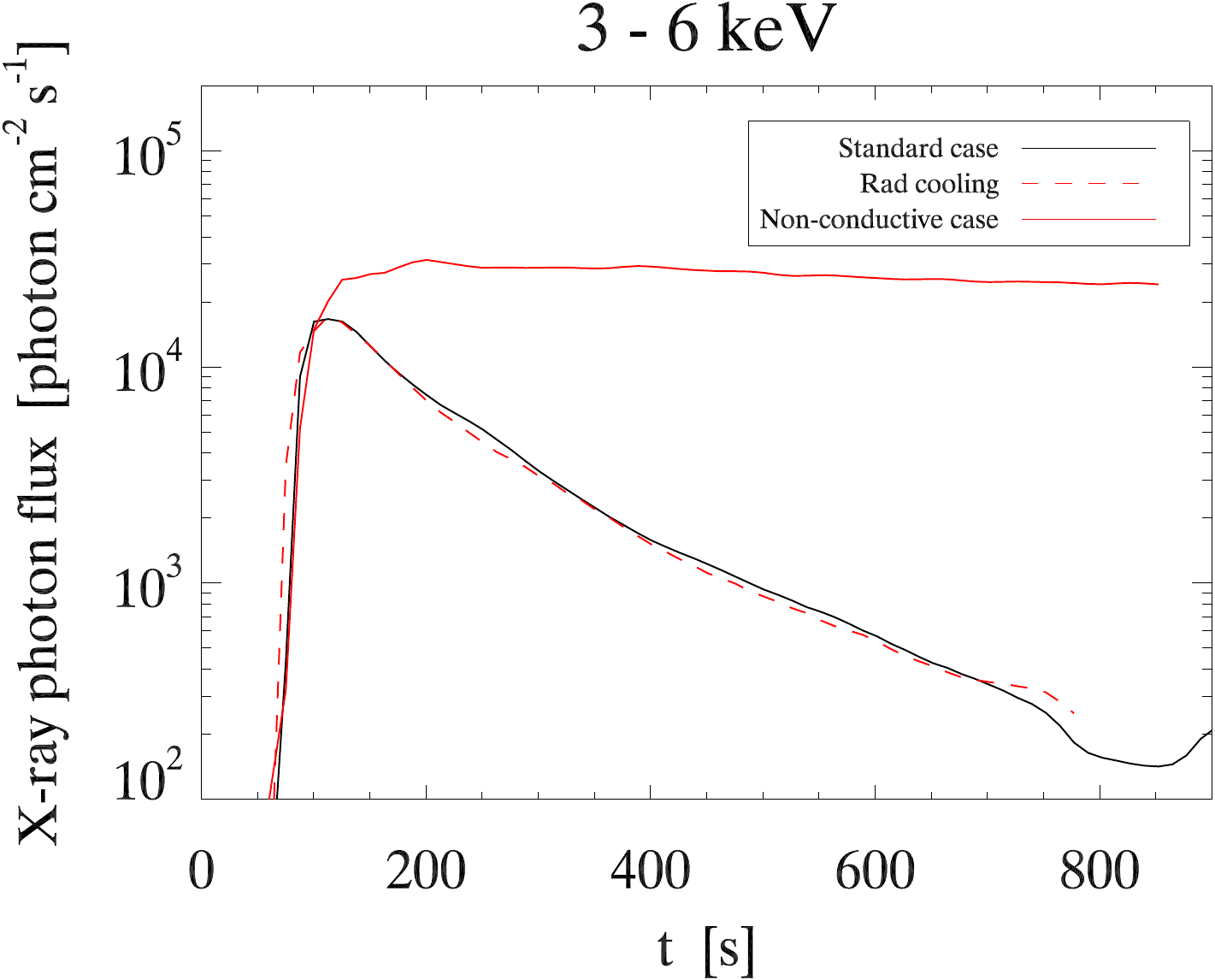} \\ \medskip
  \includegraphics[width=.95\hsize]{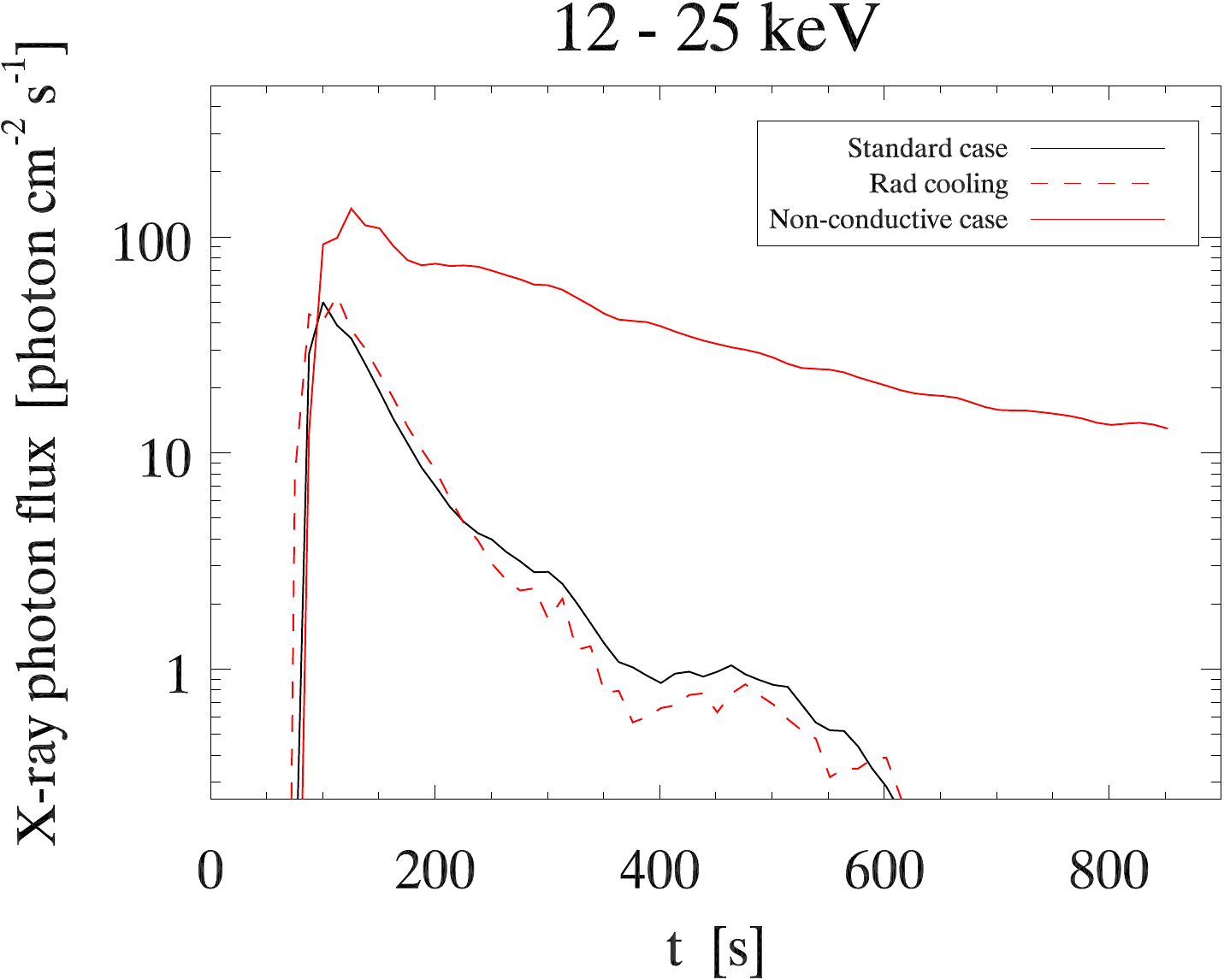} \\
  \caption{
    Light-curves for the standard case, for the standard case with radiative cooling and for the standard case without thermal conduction.
    As expected, radiative cooling as very little effect for the loop parameters chosen.
    Conductive cooling (and leakage), on the other hand, plays a very important role during the whole relaxation phase.
    Top panel: Light-curves in the $3-6\un{keV}$ range.
    Bottom panel: Light-curves in the $12-25\un{keV}$ range.
  }
  \label{fig:lightcurves_cooling}

\end{figure}

Figure \ref{fig:lightcurves_comparative} (top panel) shows a series of light-curves for different cases, all in the broad $1-25\un{keV}$ photon energy range.
The cases represented are the standard case (black line), a low-twist case (blue line), a case with a flux-rope twice as thin (continuous green line), a case with a flux-rope twice as long (dashed green line), a weak $\mathbf{B}$-field case (twice as weak; continuous orange line), and a case with strong $\mathbf{B}$-field (by a factor $2$) and a denser flux-rope (denser by a factor $4$) such that the plasma beta and characteristic Alfvén speed are maintained (dashed orange line).

Variations in flux-rope geometry and magnetic field amplitude lead to different emitted peak fluxes and to different peaking time-scales.
The maximum photon flux amplitude scales approximately linearly with the flux-rope's initial magnetic energy and initial density.
This can be seen easily by comparing the standard case (with initial magnetic energy $E_{mag}^0 \propto B_0^2$) with the cases with a flux-rope twice as long, with a flux-rope twice as thin, and with a magnetic field amplitude twice as strong (but the same initial density $n_0$).
These have initial magnetic energies which are, respectively, $2E_{mag}^0$, $1/4 E_{mag}^0$ and $4E_{mag}^0$.
The corresponding peak fluxes deviate from that of the standard case by factors, respectively, $2.0$, $1/3.67$, and $4.2$.
The case with a strong-field (twice as strong) and denser flux-rope ($4$ times as dense) shows that the peak emission depends also on the density, such that the emitted flux is proportional to $B_0^2 n_0$.
This is physically sound, as the primary source of plasma heating is the initial (free) magnetic energy, and as the resulting temperature increase is proportional to the transferred energy per particle ($T_0 \propto 1/n_0$).
The photon flux is proportional to $n_0^2$, and so the final scaling factor is also proportional to $n_0$.
The time-scales required to achieve the peak flux $\tau_{peak}$ are proportional to $r_0 / v_0$ (ratio of flux-rope radius to characteristic Alfvén speed), at least for all cases with the same transverse twist distribution.
Indeed, the case with a thin flux-rope (twice as thin) has a peaking time which is about half that of the standard case, and so does the case with a strong magnetic field (twice as strong, and so with a  characteristic Alfvén speed $v_0$ twice as high) but with same radius $r_0$.
Taking the latter case and increasing the initial density $\rho_0$ such that the initial characteristic Alfvén speed is maintained in respect to the standard case leads again to the same peaking time.
\new{%
Note that the expression $\tau_{peak} \propto r_0 / v_0 = r_0 \sqrt{\rho_0} / B_0$ remains constant if the loop's length $L_0$ varies while keeping $v_0 = \sqrt{\rho_0} / B_0$ constant (compare the black and dashed green curves on Fig. \ref{fig:lightcurves_comparative}).
}
The bottom panel in Fig. \ref{fig:lightcurves_comparative} shows the same light-curves normalised by the scaling factors discussed above.
The curves fall much closer together, showing that these scaling factors proposed work well for an extended period of the evolution of the simulated coronal structures.

For completeness we discuss two additional cases similar to the standard case (same loop parameters) but one with radiative cooling turned on and the other with thermal conduction turned off (see Sect. \ref{sec:numerical-code}).
Fig. \ref{fig:lightcurves_cooling} compares the light-curves obtained for the standard case (represented with continuous lines), for the case with radiative cooling (dotted lines), and for the case without thermal conduction (red continuous line).
The total effect of the radiative cooling remains very small during the whole period of time simulated, thus confirming that plasma cooling by radiation is negligible for the loop parameters we chose.
%

Thermal conduction (and foot-point heat leakage) has, on the other hand, a strong effect.
The initial growth phase in the latter is similar in both the conductive and non-conductive cases, but the emission peak is stronger in the latter (as the plasma reaches higher maximum temperatures without thermal conduction).
More importantly, the relaxation phase is very different in the non-conductive case, as the plasma does not cool down globally and conserves its internal energy (see Fig. \ref{fig:energies}).
Hence, the emitted photon flux does not decay in that case as would be expected in regards to observed flares.
Thermal conduction is therefore a requirement for the correct modelling of the thermal emission in solar flares.
%

\subsection{Temperature distribution and emission measures}
\label{sec:dem}

The thermal X-ray spectra displayed in Fig. \ref{fig:emission_and_spectra} show that the plasma becomes intrinsically multi-thermal after the kink instability is triggered.
A small (but strongly emitting) fraction of the flux-rope's plasma is heated up to temperatures one order of magnitude above that of the background plasma.
More high energy photons are produced, and the emission spectrum elongates into the high photon energy end.
%

We tried fitting the photon flux density in Eq. (\ref{eq:photonflux}) to the spectra we obtained from the simulations in order to find the best fit values for the flare temperature and emission measure ($T_e$ and $\mathrm{EM}_e$ hereafter).
Note that this expression assumes a uniform temperature $T_e$ for the emitting plasma (which can be seen as an effective temperature).
These fits were performed for different instants of the simulations, so as to give an indication of the temporal evolution of the fitted parameters and to allow comparisons with the original simulation data.
We found that the thermal spectra can be well approximated by the emission of a volume of plasma at uniform temperature during most of the linear phase (for $t < 80\un{s}$, roughly).
The fitted spectra remains very close to the original spectra, and $T_e \approx \langle T \rangle$ (where $\langle T \rangle$ is the volume-averaged temperature in the simulation).
The situation changes dramatically as the saturation phase approaches and magnetic reconnection starts taking place.
The ohmic heating generates a temperature distribution with a very long upper-tail and emission at higher energies suddenly becomes more important, hence hardening the spectra.
The fitted curves hardly match the original spectra during the saturation phase in the whole energy range we considered here ($1 - 25\un{keV}$).
A multi-temperature fit would probably yield more consistent results (we will not attempt such techniques in this manuscript, though).

\begin{figure}[t]
  \centering

  \includegraphics[width=.96\hsize,clip=true,trim=0 0 0 30]{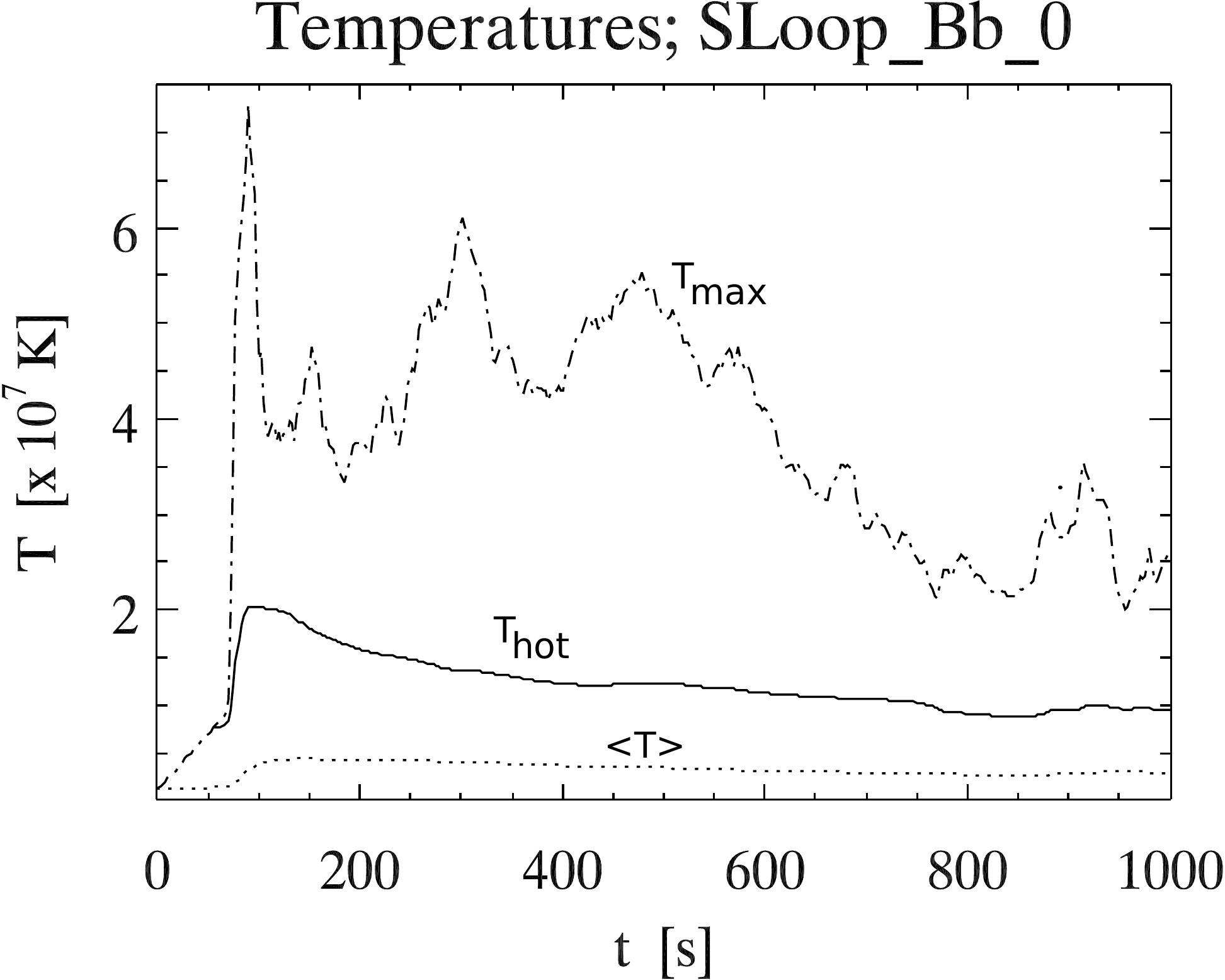} \\ \bigskip
  \includegraphics[width=.96\hsize,clip=true,trim=0 0 0 24]%
  {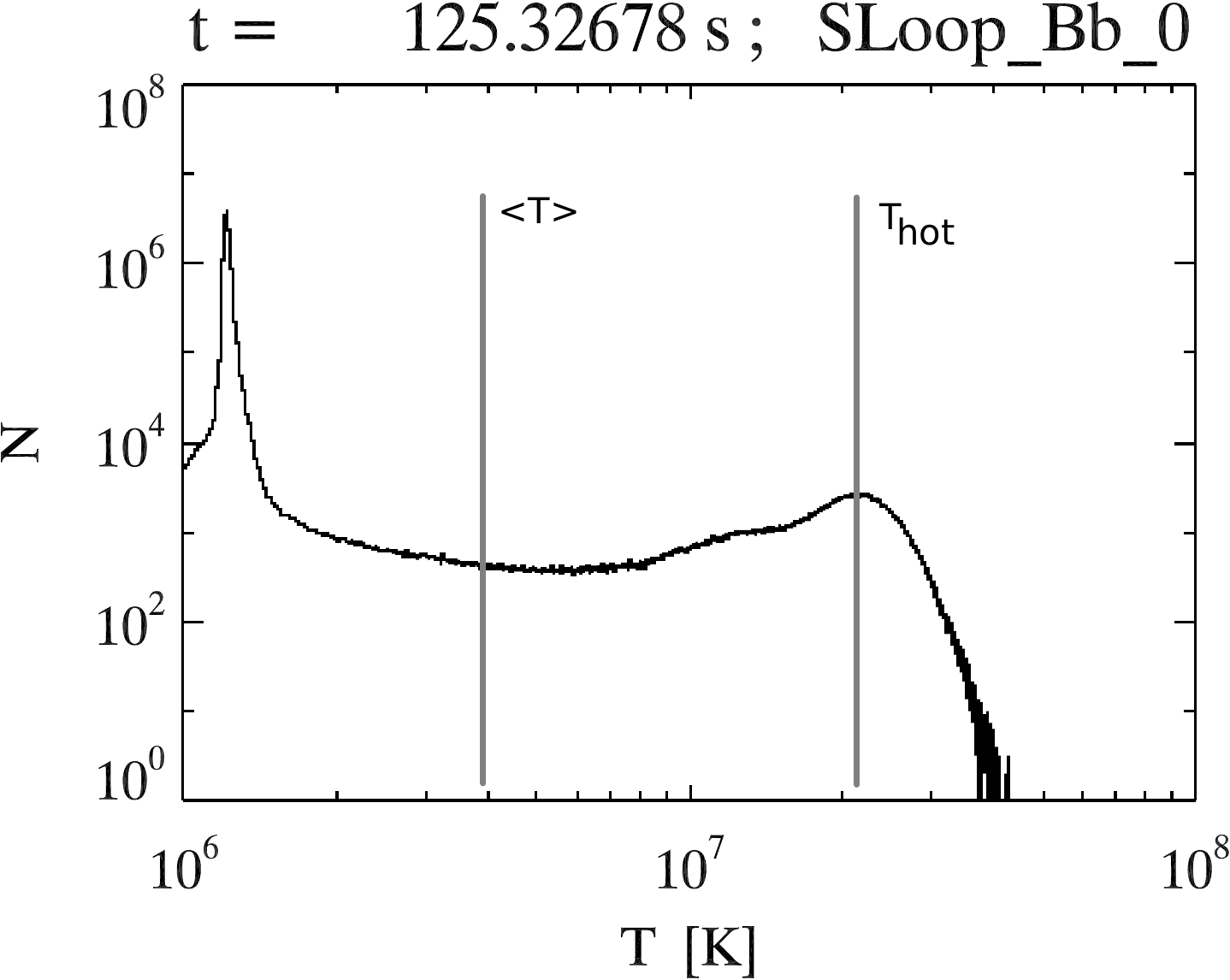}

  \caption{
    The top panel shows the temperature $T_{hot}$ corresponding to the average temperature of the bulk of the hot plasma component which develops after the saturation phase (continuous line).
    The maximum temperature $T_{max}$ and the average temperatures $\langle T \rangle$ at each instants are represented, respectively, by a dot-dashed line and a dotted line.
    The bottom panel shows a histogram of the plasma temperature in the simulation at $t=125\un{s}$.
      The vertical lines in the bottom plot mark the positions of the hot plasma component temperature and of the volume-averaged temperatures ($T_{hot}$ and $\langle T \rangle$, respectively).
  }
  \label{fig:inverse_fit}
\end{figure}

The top panel in Figure \ref{fig:inverse_fit} represents the temporal evolution of the plasma temperature in our simulations.
The maximum temperature $T_{max}$ is represented by a dot-dashed line and the volume-averaged temperature $\langle T \rangle$ is represented by a dotted line.
The continuous line represents the volume-averaged temperature of the bulk of the hot plasma component, which we will call $T_{hot}$ hereafter (providing an indication of the effective flare temperature).
The loop's plasma undergoes a strong and quick initial heating event, and cools down more slowly afterwards. 
The actual cooling time-scale is naturally larger than the estimated conductive cooling time-scale. 
Small sporadic heating events with finite duration keep occurring during the relaxation phase as the magnetic field tries to approach a potential state and contribute to maintaining the loop's plasma hotter than the background in spite of the strong conductive cooling. 
\new{The bottom panel displays a histogram of the temperature at one selected instant of the simulation ($t=125\un{s}$, just after the saturation phase) with markers identifying the values of $T_{hot}$ and $\langle T \rangle$ at that instant.}
The plasma temperature distribution develops an extended upper tail during the saturation and beginning of the relaxation phase.
It indicates the presence of both a cold plasma component and of a hot flare component.
The lower temperature peak (background plasma) is, though, a consequence of the choice of initial conditions, and broadens as the simulation proceeds.
The higher temperature peak is clearly visible during the first few minutes after the saturation, but is less pronounced and spreads outs afterwards under the action of the thermal conduction.
Overall, the temperature distribution is broad and continuous, extending across more than one order of magnitude.

\begin{figure}[!h]
  \centering
  \includegraphics[width=.855\hsize,clip=true,trim=0 55 0 24]{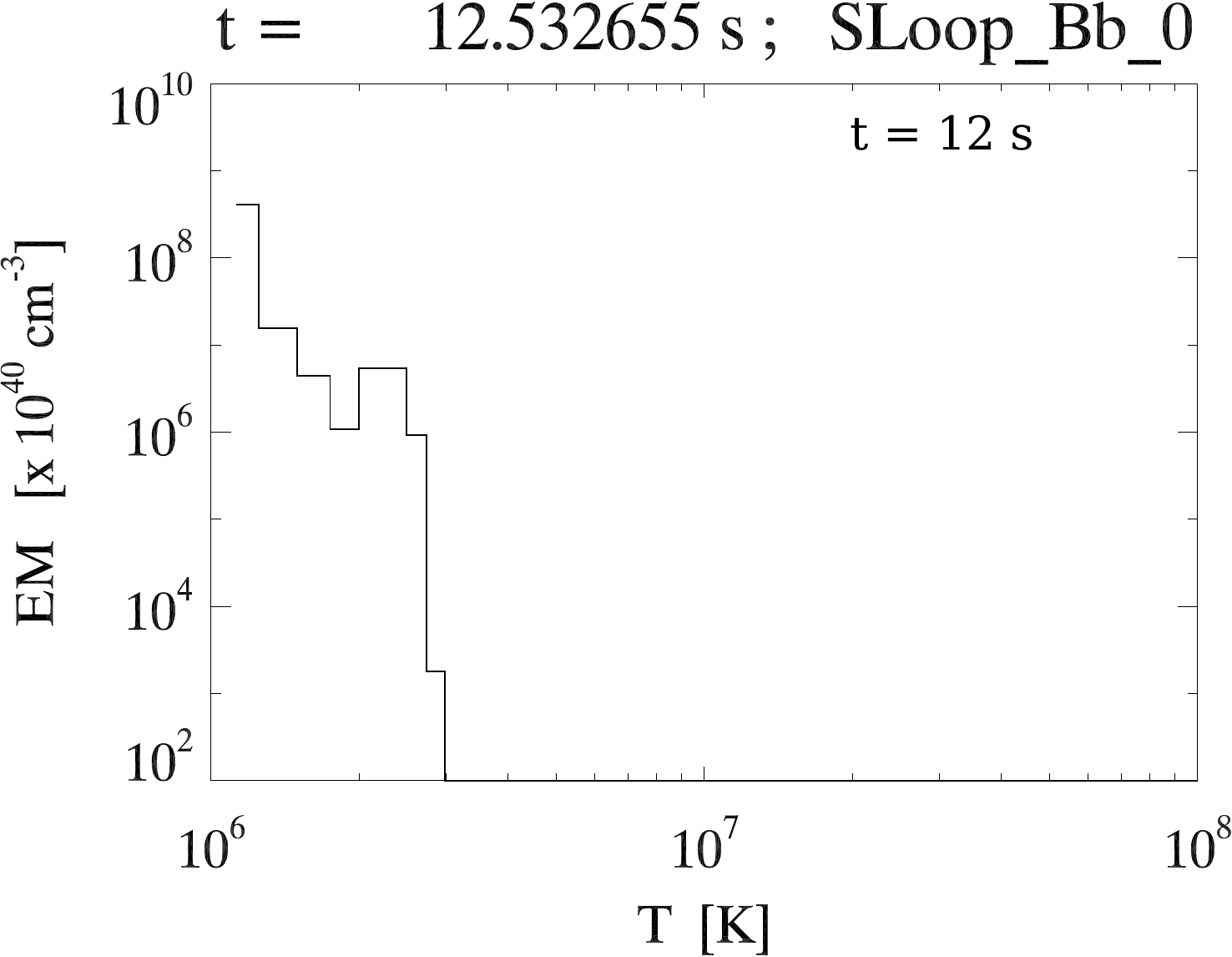} \\
  \includegraphics[width=.855\hsize,clip=true,trim=0 55 0 24]{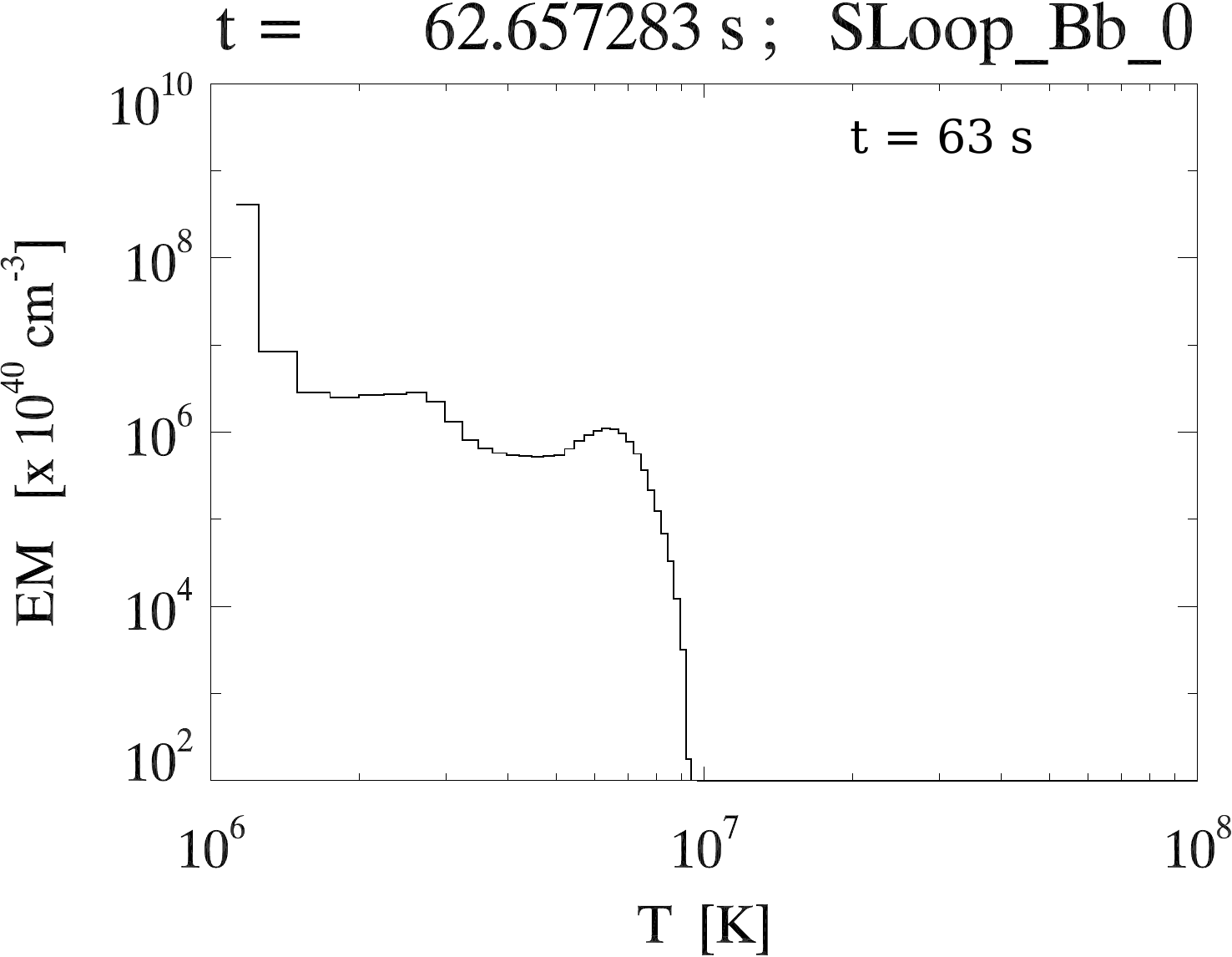} \\
  \includegraphics[width=.855\hsize,clip=true,trim=0 55 0 24]{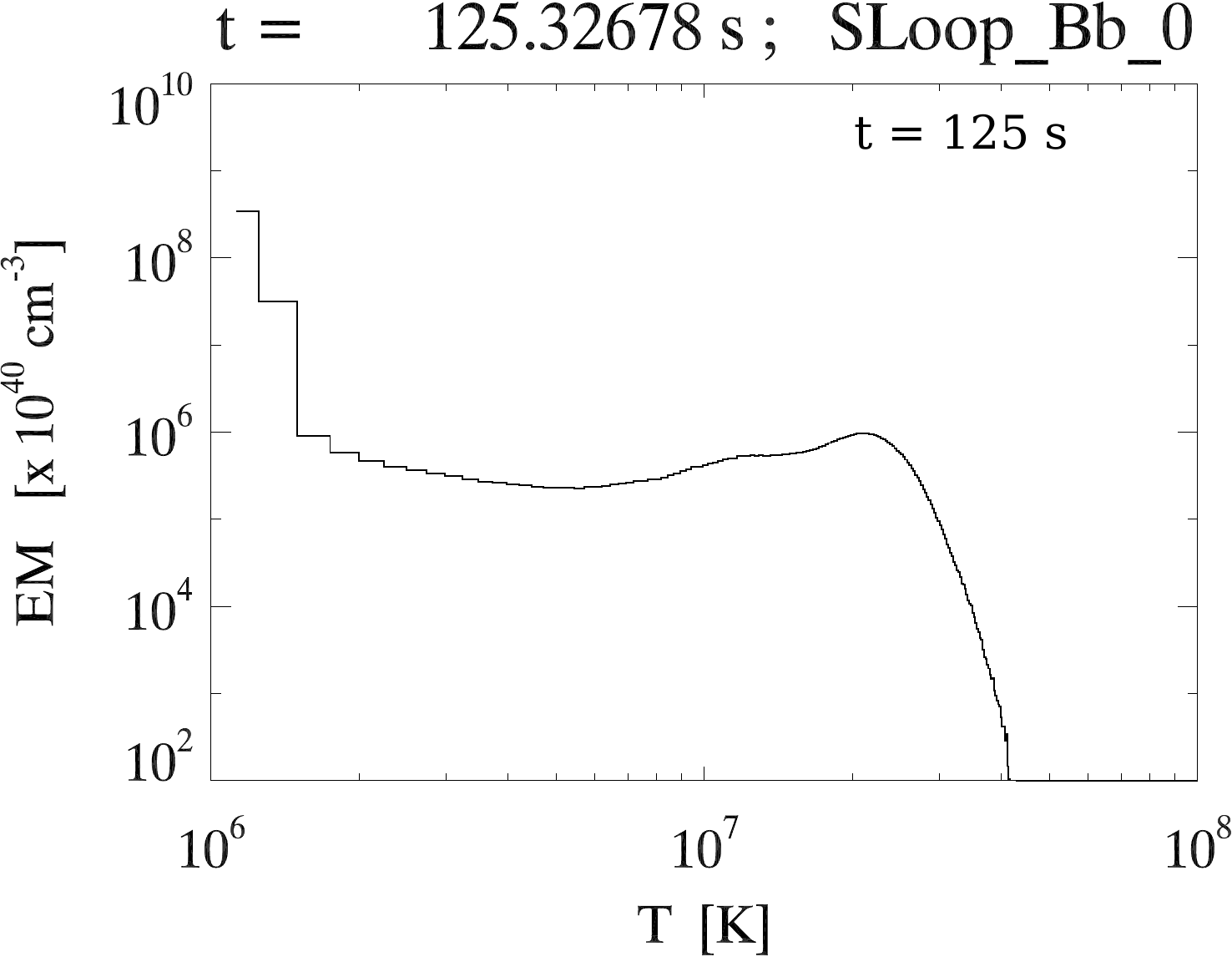} \\ 
  \includegraphics[width=.855\hsize,clip=true,trim=0 0  0 0]{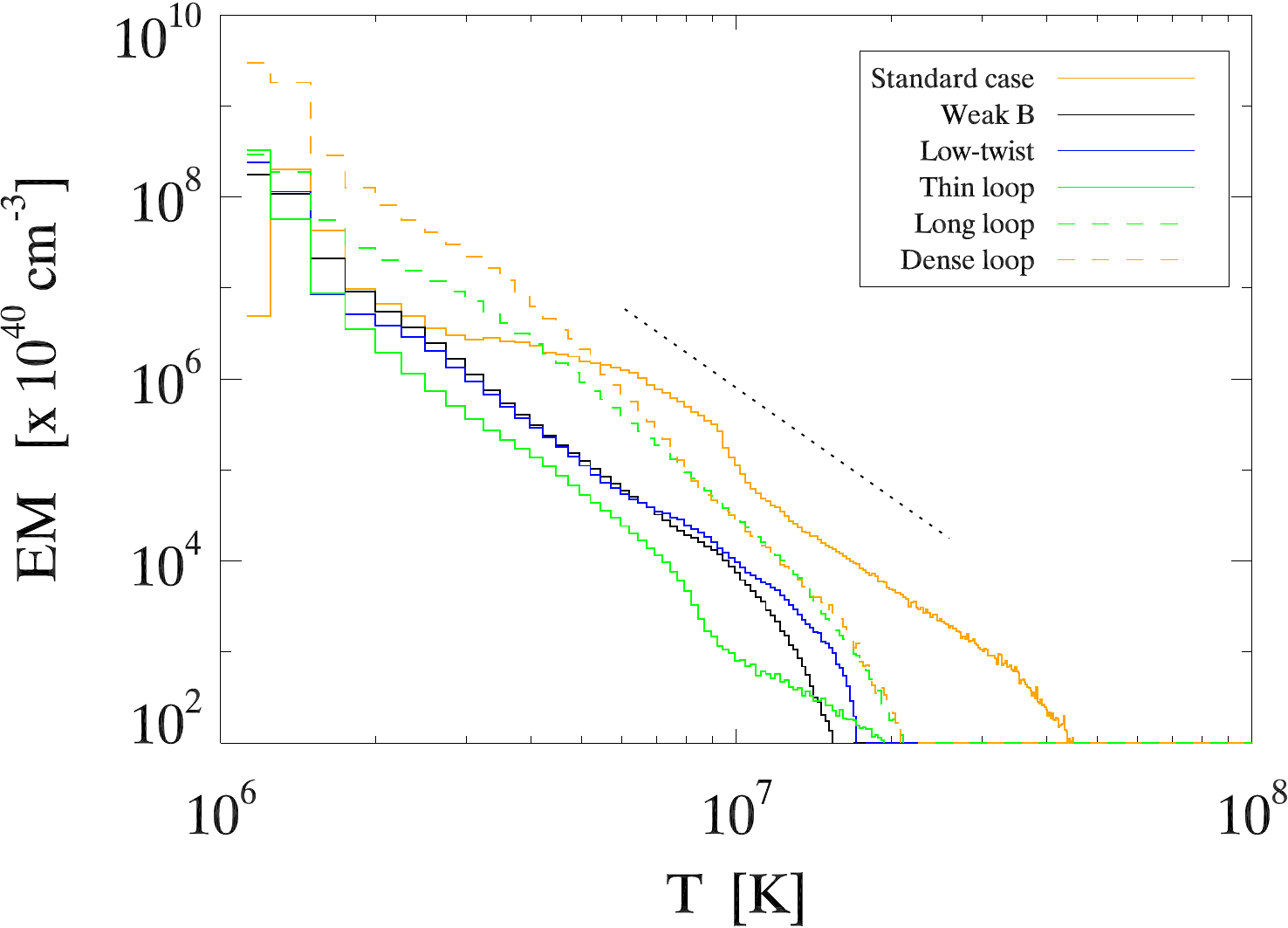}
  \caption{
    Emission measure as a function of temperature $\mathrm{EM\left(T\right)}$ at different instants, computed with temperature bins of width $\delta T = 2.5\e{5}\un{K}$.
    The instant represented in the first three panels are, from top to bottom, $t=12\un{s}$ (linear phase), $t=63\un{s}$ (beginning of the saturation phase) and  $t=125\un{s}$ (early stage of the relaxation phase).
    The initially narrow $\mathrm{EM}\left(T\right)$ (centred at $T_0 = 1.2\e{6}\un{K}$) extends quickly into the higher temperature range as the plasma is strongly heated up during the initial phases.
    The $\mathrm{EM}$ profile then slowly converges to a power-law distribution $\mathrm{EM}\propto T^{-4.2}$ for $T \gtrsim 2\e{6}\un{K}$.
    The last panel represents the relaxation phase \new{at $t \approx 25 \tau_A$ for different cases}, namely the standard case, a case with lower twist, a flux-rope twice as thin, a flux-rope twice as long, a weak-$\mathbf{B}$ and a denser flux-rope.
    The dotted line indicates the slope of a curve $\mathrm{EM}\propto T^{-4}$ for visual reference.
  }
  \label{fig:EM_temperature}
\end{figure}

\begin{figure}[!h]
  \centering
  \includegraphics[width=.855\hsize,clip=true,trim=0 0 0 24]{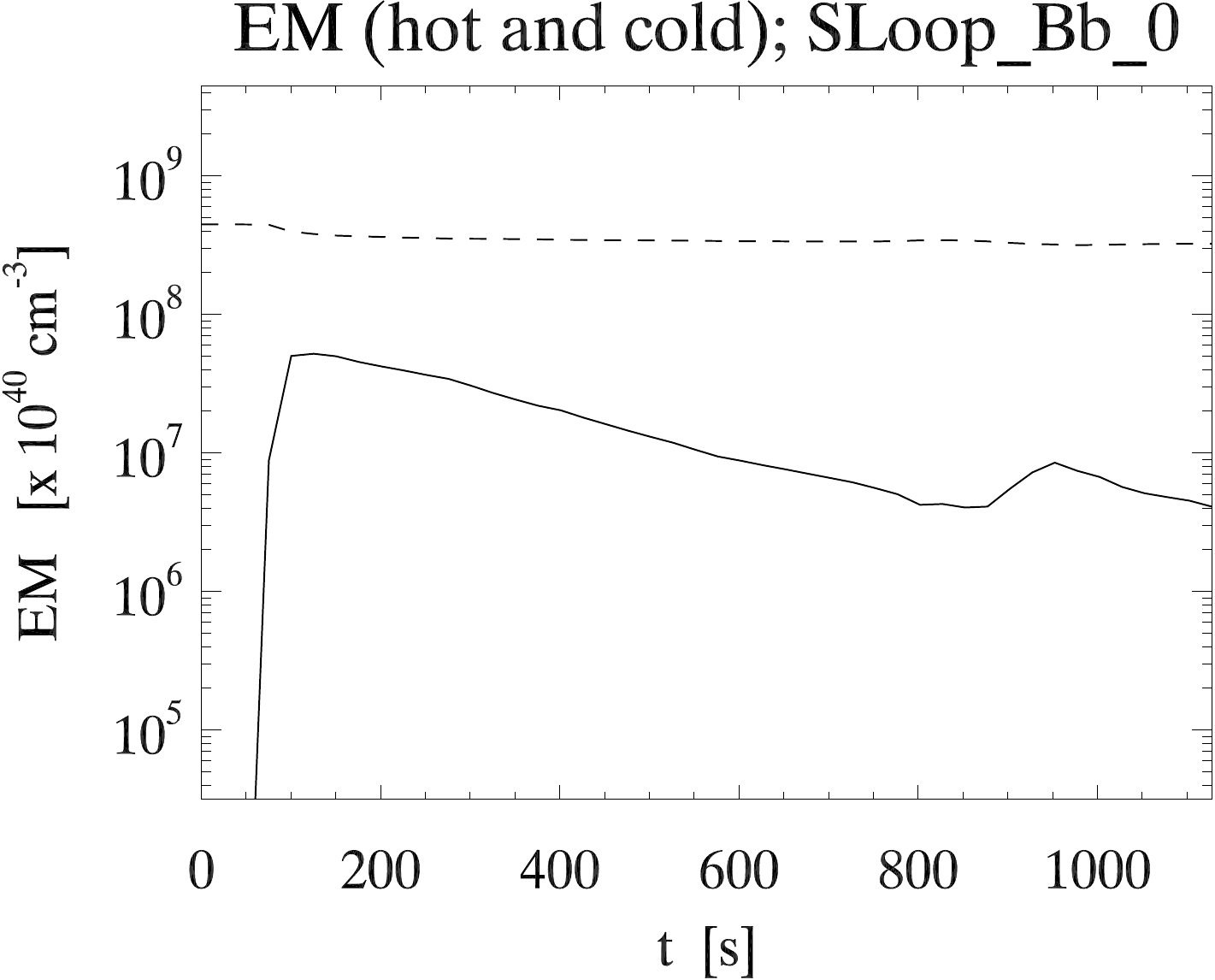}
  \caption{
    Total emission measure (EM) of the cold and of the hot plasma components as a function of time (represented, respectively, with a dashed and a continuous line).
    We define here ``hot'' all the plasma at a temperature above $9\un{MK}$ \citep[\emph{cf.}][]{sylwester_solar_2014}.
  }
  \label{fig:EM_hot_and_cold}
\end{figure}

A more interesting spectral diagnostic tool consists of considering a temperature-dependant emission measure $\mathrm{EM}\left(T\right)$ in the function $I\left(h\nu, T\right)$ (see Eq. \ref{eq:photonflux}).
\new{We computed a time-series of $\mathrm{EM}\left(T\right)$ curves directly from our simulations (see Sect. \ref{sec:methods} for a description of the method used)}.
Figure \ref{fig:EM_temperature} shows a sample of these curves for three representative instants of our standard case.
These are $t=12\un{s}$ (linear phase), $t=63\un{s}$ (start of the saturation phase) and  $t=125\un{s}$ (early stage of the relaxation phase).
Then, a series of curves corresponding to the relaxation phase for different cases are plotted together \new{(at about $t = 25 \tau_A$ for each of the cases represented)}.
These correspond to a case with lower twist, to a flux-rope twice as thin, to a flux-rope twice as long, to a strong $\mathbf{B}$ and to denser flux-rope (see the inset caption).
The dotted black line on the bottom panel of the figure is a guideline indicating the slope of the curve $\mathrm{EM}\propto T^{-4}$.
The initial $\mathrm{EM}\left(T\right)$ distribution is narrow (as expected for an isothermal plasma) and centred at $T = T_0 = 1.2\e{6}\un{K}$.
It then extends quickly into the higher temperature range, specially as the plasma is strongly heated up by ohmic diffusion during the saturation phase.
\new{A transient bump forms in the higher temperature part of the $\mathrm{EM}\left(T\right) $ distribution, in a manner which is qualitatively similar to that of the temperature distribution described above (see Fig. \ref{fig:inverse_fit}).}
The dominant plasma populations are then composed by the non-heated background plasma (\emph{i.e}, at the initial temperature $T_0$) and the strongly heated plasma.
Significant emission measure is then found for a plasma with a temperature around $20\un{MK}$.
\new{The $\mathrm{EM}$ profile will afterwards slowly converge to a power-law distribution, as the aforementioned \emph{hot} component spreads out and disappears.}
As the dotted guidelines indicate, the curve settles close to $\mathrm{EM}\propto T^{-4}$ for $T \gtrsim 2\e{6}\un{K}$.
Fitting the curve to a power-law between $T = 2\e{6}\un{K}$ and $T = 1\e{7}\un{K}$ yields 
a power-law exponent $-4.2 \pm 0.1$. 
%
We found the same behaviour in all the simulation runs we performed, with the $\mathrm{EM}\left(T\right)$ evolving in the same way and converging to a power-law with the same index (but different absolute values).
The last panel in Fig. \ref{fig:EM_temperature} shows a few illustrative cases, for which we varied the flux-rope length, thickness, level of twist and magnetic field strength.
Cases with different numerical resolutions, viscosity and magnetic resistivity were also verified.
The only exception we found was the case similar to the standard one but without thermal conduction.
In the latter, a fraction of the heated plasma reaches maximum temperatures higher by a factor $\sim 5$ \citep[\emph{cf.}][]{botha_thermal_2011}, and remains hot in the lack of a cooling mechanism as efficient as the Spitzer-Härm thermal conduction.
This naturally translates into a $\mathrm{EM}$ distribution extending up to higher temperatures, and to a flatter power-law.

Figure \ref{fig:EM_hot_and_cold} shows the temporal evolution of the total emission measure of the ``hot'' and ``cold'' plasma components separately.
We define here hot (cold) plasma here as all plasma with a temperature above (below) a threshold of $9\un{MK}$, as in \citet{sylwester_solar_2014}.
The emission measure of the hot plasma component increases abruptly during the linear/impulsive phase of the kink instability, reaching a maximum value of $5\e{47}\un{cm^{-3}}$ in $\sim 125\un{s}$.
It then slowly decays during the relaxation phase.
This translates into an inverse variation with the same absolute amplitude the evolution of the total emission measure of the cold component (the relative amplitude is much smaller, though).


\section{Discussion}
\label{sec:discussion}


We studied the properties of the thermal continuum X-ray emission in kink-unstable coronal loops by means of numerical MHD simulations. 
The model we used consists of twisted magnetic flux-ropes embedded in a uniform background coronal field.
Given that the magnetic twist is strong enough, the development of the kink instability provides a viable mechanism for the liberation of big amounts of free magnetic energy initially stored in the twisted field, as demonstrated by many previous studies.
The numerical simulations presented here aim at providing a good description of its thermodynamics and thermal X-ray emission properties, as they evolve in time off from their initial highly-twisted and quasi-stationary state.
For this purpose, it was imperative to consider a set of compressible MHD equations with viscous, resistive and conductive effects taken into account self-consistently.
Variations of plasma density and temperature reflect the dynamics and the heat transfers occurring in the system after the triggering of the kink instability. 
\new{These translate into variations of the continuum X-ray emissivity (note that the emissivity depends strongly on temperature, but also on density; see Eqs. \ref{eq:emissivity} and \ref{eq:photonflux})}.

\subsection{Comparison with SXR observations}

\new{Let us now discuss the results described in this manuscript with respect to observations of solar flares in soft X-rays.}
Despite of the simplicity of the underlying model, a few interesting conclusions can be drawn from our simulations.

As shown in Figs. \ref{fig:emission_and_spectra} and \ref{fig:loop_closeup} (and described in Sect. \ref{sec:thermal-emission}), the thermal X-ray emission starts by appearing near the axis of the flux-rope assuming a helical and filamentary shape.
It then fills up all the flux-rope volume (during the peak of emission), and later fades away progressively during the relaxation phase.
If our model represented a real solar flare, these details would be detectable only as a fast initial increase in thickness and length of the flaring coronal loop \citep[\emph{cf.}][]{jeffrey_temporal_2013}, given the spatial resolution of the current X-ray instruments.
Interestingly, the initial magnetic twist in the pre-flare flux-rope as perceived from the X-ray emission is much smaller than the maximum twist at those instants.
This effect would also not be detected by X-ray instruments (due to spatial resolution constraints), but would probably be under the reach of the current EUV observations.
This partly is due to the geometry of the emission patterns in the initial phases of the simulated flare.
Different flux-rope twist profiles and coronal loop global geometries could perhaps lead to a different scenario, thus requiring further investigation to assess whether this result is general or specific to our model.
In any case, the flaring loop has already lost a significant fraction of its initial twist when it becomes visible.

As shown in Fig. \ref{fig:lightcurves}, the emission light-curves show an impulsive initial growth followed by a slower decay.
\new{The decay phase is faster the higher the photon energy is (and slower the lowest the photon energy is), as is the case for the solar flare light-curves measured by RHESSI and GOES in \citet{sylwester_solar_2014} for the $1-8\un{\AA}$, $0.5 - 1 \un{\AA}$, $6-12\un{keV}$ and $12-25\un{keV}$ bands.}
  \new{Fig. \ref{fig:lightcurves} shows that the maximum X-ray flux obtained is of about} $1.2\e{5}\un{photons\ cm^{-2} s^{-1}}$ in the $1-3\un{keV}$ band, $2\e{4}\un{photons\ cm^{-2} s^{-1}}$ in the $3-6\un{keV}$ band, $2\e{3}\un{photons\ cm^{-2} s^{-1}}$ in the $6-12\un{keV}$ band and $50\un{photons\ cm^{-2} s^{-1}}$ in the $12-25\un{keV}$ band.
\new{The predicted fluxes are consistently higher than the thermal emission of the B class flares detected by RHESSI discussed by \citet{hannah_rhessi_2008}, and fall closer to fluxes typical of a C class flare.}

\new{The simulated X-ray spectra are strongly multi-thermal (see the spectra in Fig. \ref{fig:emission_and_spectra} and the temperature distribution in Fig. \ref{fig:inverse_fit}).
This is due to the strong ohmic heating occurring during the saturation phase, and translates into a broad distribution of the $\mathrm{EM}$ as a function of temperature (see Fig. \ref{fig:EM_temperature}).}
\new{The $\mathrm{EM}\left(T\right)$ distributions we obtained clearly show two distinct components (see Fig. \ref{fig:EM_temperature}).
The low-temperature component is centred at the initial temperature in our model (slightly above $10^6{K}$), and represents the background coronal plasma temperature.
The high-temperature component corresponds to the fraction of the plasma impulsively heated during the saturation phase (in a time-scale of the order of $150\un{s}$).
The actual temperature of this hot component (corresponding to $T_{hot}$ in Fig. \ref{fig:inverse_fit}) depends on the exact parameters of the simulated coronal loops.
Cases with stronger magnetic fields and/or lower densities will reach higher $T_{hot}$ values than cases with weaker magnetic fields and/or higher densities.
The minimum and maximum values of $T_{hot}$ we obtained for the parameter range we explored were, respectively, $5\un{MK}$ and $30\un{MK}$. 
Our standard case reached $T_{hot} \approx 20\un{MK}$ (see Fig. \ref{fig:inverse_fit}).}

These results share some similarities with the $\mathrm{EM}\left(T\right)$ distributions deduced from recent flare observations.
\new{This quantity is accessible to observers by comparing photon flux measurements at different energy bands. 
This is a subject of active research in the field of the extreme ultra-violet wavelengths \citep[\emph{e.g}][]{aschwanden_automated_2013,hannah_differential_2012}, but much less is known about the temperature dependence of the emission measure in the soft X-ray range \citep{reale_evidence_2009,battaglia_rhessi_2012}.}
Using combined RHESSI and SDO/AIA data, \citet{battaglia_rhessi_2012} found that the $\mathrm{EM}$ distribution of flaring loop's plasma they studied had two temperature components, one at around $2\un{MK}$ and one at or around $8\un{MK}$.
Furthermore, the hot component became progressively more preponderant as the flare proceeded, while the cold component remained fairly unchanged during the same period of time.
\new{As in our simulations, they interpreted this feature as the contributions of the hot flare plasma (the \emph{hot} component) and of the background coronal plasma (the \emph{cold} component)}.
%
%
\citet{sylwester_solar_2014} have also shown similar results using RESIK data for a GOES class $M1.0$ flare showing a clear two-temperature structure during the peak phase.
The colder plasma had an approximately constant temperature of about $3 - 6\un{MK}$ and the hotter plasma a temperature in the range of $16 - 21\un{MK}$.
\citet{prato_regularized_2006} have also found RHESSI spectra consistent with approximately isothermal plasma components at low temperature and  very broad forms of the EM at high temperatures.

\new{
During the relaxation phase, as the simulated flux-ropes relax towards a much lower-twist state, the $\mathrm{EM}$ converges asymptotically to a power-law $\mathrm{EM} \propto T^{-4.2}$ (see Fig. \ref{fig:EM_temperature}).
It would be interesting to verify this result observationally, although the X-ray emissivity drops to very low values during this late phase (possibly below the detection threshold), the actual coronal plasma might be perturbed by other events, and additional physical processes could also come into play during these long post-flare time-scales (such as radiative cooling and mass loading processes).
In fact, in most of the $\mathrm{EM}$ measurements cited above, the high-temperature tail of the EM distributions is considerably steeper than that of the \emph{asymptotic} limit $\mathrm{EM} \propto T^{-4.2}$ we propose here.}
The exception is perhaps that of the ``region 2'' in \citet{battaglia_rhessi_2012}.
The $\mathrm{EM}$ distributions of this region particularly resembles the ones we calculated (Fig. \ref{fig:EM_temperature}).
\new{In this region, the \emph{hot} component first grows in amplitude, and then spreads out into a flatter power-law like distribution, reaching temperatures above $30\un{MK}$.}
The steeper $\mathrm{EM}$ falloff at the high end of the distributions obtained in most observations is better represented in our simulations by the upper tail of the \emph{hot} component, as it evolves from the saturation phase ahead (\emph{i.e}, in the second and third panels in Fig. \ref{fig:EM_temperature}).
The equivalent power-law index of the latter (to the right of the ``bump'' in Fig. \ref{fig:EM_temperature}) varies between $-9$ and $-6$, closer to the observed values.
Our simulations thus suggest that the steep $\mathrm{EM}\left(T\right)$ falloff in the high-temperature range is related to the transient heating phenomena which immediately follow a flare.

\new{Figure \ref{fig:EM_hot_and_cold} shows the temporal evolution of the total emission measure of the hot flare plasma (defined as having a temperature above $9\un{MK}$) in our simulations.
At the peak of emission, the emission measure of the hot plasma reaches a value of the order of $5\e{47}\un{cm^{-3}}$, which is, for example, one order of magnitude below what was measured by \citet{sylwester_solar_2014} for an $M$ class flare.}

\subsection{Scope and caveats of the model}
\label{sec:caveats}

\new{The origin of the twisted magnetic flux-ropes in the corona is most probably related to a combination of flux-emergence, magnetic shearing by surface motions and magnetic reconnection in the corona, but the exact details are unknown at the present date \citep{jouve_three-dimensional_2009,fan_origin_2009,jouve_global_2013,pinto_flux_2013}.
We only consider here the dynamics of already existing coronal flux-ropes\new{, already at typical coronal background temperatures}. 
The study of their generation is out of the scope of this manuscript\new{, as is the general problematic of the heating of the coronal loops}.
Furthermore, we consider only the coronal part of such magnetic loops and set up boundary conditions which are meant to represent the effects of the dense and cold sub-coronal layers.
The magnetic field is line-tied to the top and bottom boundaries (which remain stationary), and heat is allowed to be conducted outwards in order to let the loop cool down conductively (see Sect. \ref{sec:numerical-code}).
It is worth noting that even though a finite heat flux across the foot-points is allowed, its magnitude could possibly be underestimated in respect to the heat flux from the real corona onto the much colder chromosphere.
}
The line-tying condition is widely used in this kind of study, being thought of as a proxy to the way the corona reacts quickly (in a time-scale of the order of the Alfvén crossing time) to the much slower surface dynamics.
It must be noted, nevertheless, that this assumption may overestimate the amount of magnetic energy in coronal loops in time-scales longer than a few Alfvén crossing times \citep{grappin_mhd_2008}.
To verify the severity of this issue, we performed additional runs of our standard case with different top and bottom boundary conditions (open and periodic).
We found that the dynamical evolution of the system was nearly unaffected during the linear (impulsive) and saturation phases.
The magnetic field relaxation is faster, though, if magnetic and kinetic energy are allowed to flow outwards through the footpoints.
The magnetic field-lines approach their final state faster, show less low amplitude oscillations and may loose all their helicity.
But, more importantly to the outcome of this paper, the overall thermal behaviour of the system (heating and emission patterns) is maintained.
Note that most of the plasma heating is due to local ohmic dissipation following the initial impulsive kinking phase (as opposed to the viscous dissipation of the flows, for example).
\new{
Not including the chromospheric layers in the numerical domain furthermore means that mass transfer between the corona and the chromosphere is not taken into account.
It is well known, nevertheless, that thermal conduction and electron collisions may heat up the dense plasma near the loop's footpoints and hence cause chromospheric evaporation during the course of a flare \citep{acton_chromospheric_1982,antonucci_energetics_1984,yokoyama_two-dimensional_1998,yokoyama_magnetohydrodynamic_2001,benz_flare_2008}, and that subsequent mass draining can occur.
Observations by \citet{mckenzie_electron_1980} and \citet[][]{saint-hilaire_statistically_2010} show that the plasma upflows due to chromospheric evaporation can raise the flare plasma densities up to $1\e{11}\un{cm^{-3}}$ at coronal heights.
This can be important in the assessment of the overall energy budget of a flaring system, as the radiative cooling efficiency and thermal emission depend strongly on plasma density.
}
Present day numerical models including (at least part of) the transition region and chromosphere\new{, however, show only a modest degree} of chromospheric evaporation following the onset of the kink instability in coronal loops, and during the typical dynamical time-scales covered by this type of study (Gordovskyy, Browning, private communication)\new{,
perhaps for the reasons discussed in \citet{bradshaw_influence_2013}}.
Future work could help on the full assessment of the effects of mass and energy transfer between the chromospheric and the coronal layers, but such an exercise is beyond the scope of this manuscript.

%
%

\new{
Estimations of the properties of the thermal X-ray emission are made here as a post-processing step, which means that the associated energy losses are not self-consistently accounted for in the MHD simulations.
However, the total energy radiated away in this energy range is negligible in respect to the plasma's thermal energy (see Sect. \ref{sec:model-parameters} and \ref{sec:thermal-emission} for related discussions).
}
\new{Besides, our study covers a set of parameters for which plasma cooling is strongly dominated by conductive losses rather than by radiative losses, and for which the characteristic cooling time-scale of the latter is longer than the dynamical time-scales we are dealing with. 
Substantially denser and/or colder coronal loops could, however, require radiative losses to be taken into account self-consistently (\emph{n.b}, the ratio of conductive to radiative cooling rates is proportional to $n^2 L^2 / T^4$).
We verified \emph{a posteriori} that the assumption of conductive cooling regime is correct for the cases studied here(see Fig. \ref{fig:lightcurves_cooling}).}
\new{For simplicity, we consider only the thermal continuum emission in the simulated flaring loops, and do not take into account X-ray line emission.
It should nevertheless be noted that line emission can be important at low photon energies, superposing to the soft X-ray spectra \citep{mckenzie_electron_1980,phillips_solar_1982,reale_sun_2001}, but without contributing with a lot of flux in the wide energy bands considered here.
Non-thermal emission can be significant at the high end of the photon energy range considered here \citep[e.g,][]{krucker_hard_2008}, but its study are besides the aims of the current manuscript.
In any case, the computed emission measures ($\mathrm{EM}$) are not affected by these simplifications as they only depend on the density distribution in the simulated volume of plasma given directly by the MHD simulations (see the definitions in Sect. \ref{sec:emission} and the discussion in Sect. \ref{sec:dem}).}
%


\section{Summary}
\label{sec:summary}

%
%

\new{
  In this paper, we investigate the properties of the thermal continuum X-ray emission produced in kink-unstable magnetic flux-ropes by means of numerical MHD simulations.
  The model consists of a kink-unstable twisted magnetic flux-rope embedded in a uniform coronal background field \citep[as in, \emph{e.g},][]{hood_coronal_2009,botha_thermal_2011,gordovskyy_magnetic_2011}.
  The system is initially at coronal temperatures (typically $1.2\un{MK}$), but the flux-rope plasma ends up reaching temperatures as high as $30\un{MK}$ following the triggering of the kink-instability.
  We analyse the variations of the plasma density and temperature in order to estimate thermal (continuum) emission in the soft X-ray range, as well as the emission measure distributions $\mathrm{EM\left(T\right)}$ (see Sect. \ref{sec:emission}).
  The (strong) density variations are due to plasma compression only, as we do not take into account mass transfer between the corona and the chromosphere (hence leaving out the effects of chromospheric evaporation on the density structure of the loops).

  The system undergoes three distinct phases: a linear phase during which the kink instability is triggered and grows linearly, a saturation phase during which a strong reconnexion event occurs accompanied by a strong enhancement in ohmic heating, and a relaxation phase during which the loop approaches its minimal energy state and cools down globally.
  During the initial (linear) phase of the instability, moderate plasma heating occurs due to compression (as the kinking motions of the flux-rope grow in amplitude).
  Ohmic diffusion then takes over as the instability saturates, provoking a strong and quick heating event. The flux-rope plasma is, as a consequence, heated up to temperatures between $10$ and $30\un{MK}$ (see Fig. \ref{fig:inverse_fit}). 
  Correspondingly, a ``hot'' plasma component becomes readily visible in the $\mathrm{EM\left(T\right)}$ distributions in the same temperature interval (see Fig. \ref{fig:EM_temperature}).
  Overall, significant emission measures arise for plasma at temperatures higher than $9\un{MK}$ during the peak/saturation phase (see Fig. \ref{fig:EM_hot_and_cold}).
  This type of behaviour is in agreement with measurements of emission measures in solar flares  \citep[\emph{e.g}][]{sylwester_solar_2014}.
  The X-ray emission is quickly enhanced during the saturation phase (see Fig. \ref{fig:lightcurves}) and the thermal X-ray spectrum becomes harder and clearly of a multi-thermal nature (see Figs. \ref{fig:emission_and_spectra} and \ref{fig:EM_temperature}).
  The magnetic flux-rope then relaxes progressively towards a lower energy state as it reconnects with the background flux.
  The loop plasma keeps suffering small sporadic heating events, but cools down globally by thermal conduction.
  During this phase, the thermal X-ray emission concentrates into field-aligned filaments and fades away progressively.
  The ``hot'' component of emission measure distribution spreads out slowly and converges to the power-law distribution $\mathrm{EM}\propto T^{-4.2}$.

  Overall, the amount of twist perceived directly from the continuum emission patterns is substantially smaller than the maximum twist in the simulated flux-ropes (by at least a factor $2$; see Figs. \ref{fig:loop_closeup} and \ref{fig:loop_closeup_blines}).
  Individual field-lines are clearly visible only during the late phase of the instability, after the flux-rope has already lost most of its twist.
  During the saturation phase, when the emission flux is at its maximum, the emission pattern traces the large scale displacements of the flux-rope's axis (writhe) rather than the actual twist of the magnetic field-lines.
  This effect is stronger if the spatial resolution and dynamical range are lowered in order to match those achievable by current and future X-ray instruments (as the details of the fine structure are lost).
  This result suggests that the observed lack of sufficient twist (\emph{i.e} flux-ropes twisted above the kink-instability threshold are very rarely observed) does not invalidate the kink-instability scenario for confined flares.
}

%
%

\new{%
Future work should consider the effects on the chromospheric layers on the evolution of these systems, in order to characterise more precisely the downward conductive heat flux and consequent plasma evaporation.
Particle acceleration in the reconnection sites should be taken into account in order to provide a combined view of the non-thermal and thermal X-ray emission during a flare.}
Different (more realistic) magnetic configurations should be tested and compared, either by introducing different twist profiles and more complex global loop geometries.

\begin{acknowledgements}
  This work was supported by the French Space Agency (CNES) and used computational facilities from the IDRIS and the TGCC-CEA (GENCI project 1623).
  We thank the PNST programme and P. Browning, M. Gordovskyy, O. Limousin, A. Meuris and K. Shibata for fruitful discussions.
  We acknowledge A. Mignogne and colleagues for the active development and maintenance of the PLUTO code.
  \new{We thank the anonymous referee for his suggestions, which lead to major improvements to this manuscript.}
\end{acknowledgements}

   \bibliographystyle{aa}
   \bibliography{/data/BIBTEX/refs}


\end{document}